\documentclass[12pt]{article}
\usepackage{amssymb}
\usepackage{amsmath}
\usepackage{latexsym}
\usepackage{epsfig}
\usepackage{graphicx}
\usepackage{subfigure}
\usepackage{wasysym}
\usepackage{threeparttable}
\usepackage{natbib}
\usepackage{color}
\usepackage{epstopdf}

\newcommand{\mbf}[1]{\mathbf{#1}}

\usepackage{hyperref}

\def\RR{\mathbb R}

\def\EE{\mathbb E}

\def\CC{\mathbb C}

\def\tends{\rightarrow}

\newtheorem{Proposition}{Proposition}

\newcommand{\Appendix}
{%\appendix
\def\thesection{Appendix~\Alph{section}}
\def\thesubsection{A.\arabic{subsection}}
}

\def\log{\hbox{log}}

\def\boxit#1{\vbox{\hrule\hbox{\vrule\kern6pt
          \vbox{\kern6pt#1\kern6pt}\kern6pt\vrule}\hrule}}

\def\bse{\begin{eqnarray*}}
\def\ese{\end{eqnarray*}}
\def\be{\begin{eqnarray}}
\def\ee{\end{eqnarray}}
\def\bq{\begin{equation}}
\def\eq{\end{equation}}
\def\bse{\begin{eqnarray*}}
\def\ese{\end{eqnarray*}}

\begin{document}
\thispagestyle{empty} \baselineskip=28pt

\begin{center}
{\LARGE{\bf The Cepstral Model for Multivariate Time Series: The Vector Exponential Model}}
\end{center}

\baselineskip=12pt

%%
%%
%%
%%%%%%%%%%%%%%%%%%%%%%%%%%%%%%%%%%%%%%%%%%%%%%%%%%%%%%%%%%%%%%%%%%%%%%%%
\vskip 2mm
\begin{center}
Scott H. Holan\footnote{(\baselineskip=10pt to whom correspondence should be addressed)
Department of Statistics, University of Missouri,
146 Middlebush Hall, Columbia, MO 65211-6100, holans@missouri.edu},
Tucker S. McElroy\footnote{\baselineskip=10pt Center for Statistical Research and Methodology, U.S. Census Bureau, 4600 Silver Hill Road, Washington, D.C. 20233-9100, tucker.s.mcelroy@census.gov},  and Guohui Wu\footnote{\baselineskip=10pt Department of Statistics, University of Missouri, 146 Middlebush Hall, Columbia, MO 65211-6100, gwgc5@mail.missouri.edu}\,\footnote{\baselineskip=10pt This report is released to inform interested parties of research and to encourage discussion.  The views expressed on
statistical issues are those of the authors and not necessarily
those of the U.S. Census Bureau.}  
\\
\end{center}
%
%
%
%
%%%%%%%%%%%%%%%%%%%%%%%%%%%%%%%%%%%%%%%%%%%%%%%%%%%%%%%%%%%%%%%%%%%%%%%%
\vskip 4mm

\begin{center}
{\bf Abstract}
\end{center}
Vector autoregressive (VAR) models have become a
staple in the analysis of multivariate time series and are
formulated in the time domain as difference equations, with an
implied covariance structure.  In many contexts, it is
desirable to work with a stable, or at least stationary,
representation.  To fit such models, one must impose
restrictions on the coefficient matrices to ensure that certain
determinants are nonzero; which, except in special cases, may
prove burdensome.  To circumvent these difficulties, we propose
a flexible frequency domain model expressed in terms of the spectral
density matrix.  Specifically, this paper treats the modeling
of covariance stationary vector-valued (i.e., multivariate)
time series via an extension of the exponential model for the
spectrum of a scalar time series.  We discuss the modeling
advantages of the vector exponential model and its
computational facets, such as how to obtain Wold coefficients
from given cepstral coefficients.  Finally, we demonstrate the utility of our
approach through simulation as well as two illustrative data examples focusing on multi-step ahead forecasting and estimation of squared coherence.

\baselineskip=12pt

%%%%%%%%%%%%%%%%%%%%%%%%%%%%%%%%%%%%%%%%%%%%%%%%%%%%%%%%%%%%%%%%%%%%%%%%
%
%
%

\baselineskip=12pt
\par\vfill\noindent
{\bf Keywords:}  Autocovariance matrix; Bayesian estimation;
Cepstral; Coherence; Spectral density matrix; Stochastic search variable selection; Wold
coefficients.
\par\medskip\noindent
\clearpage\pagebreak\newpage \pagenumbering{arabic}
\baselineskip=24pt
\section{Introduction}
This paper treats the modeling of covariance stationary
vector-valued (i.e., multivariate) time series through an extension
of the exponential model of \citet{Bloom:1973}.  Such a process
will be called VEXP, for Vector EXPonential. In contrast to VAR and VARMA models,
the VEXP processes that we define herein are always invertible, which means that the (causal) Wold form of the process can be inverted
into a (stable) VAR form -- or equivalently, that the spectral
density matrix of the VEXP is non-singular at all frequencies.
Necessarily, a VEXP process is also stable, or stationary, which
here means that the spectral density matrix has finite determinant
at all frequencies.  We note that, when
estimation proceeds in an unconstrained fashion (e.g., by ordinary least squares) a VAR or VARMA process need not
be stable and invertible; see \citet{lutkepohl2007new} for a basic treatment.
Nevertheless, there are practical scenarios where these restrictions
on the vector process are actually necessary.

Although Gaussian maximum likelihood estimation can still
proceed when a vector process is non-invertible, so long as the
singularities occur at a set of frequencies that have Lebesgue
measure zero (see \citet{mcelroy2012signal} for proof and
discussion), Whittle estimation (i.e., the Gaussian
quasi-maximum likelihood procedure described in \citet{taniguchi2000asymptotic})
 becomes intractable.  In addition, the long-term forecasting filters are not well-defined (see the discussion in
\citet{mcelroy2012multi}), because such filters
rely on the ability to recover the innovations from the Wold
form of the process.  Another motivation for using invertible
processes arises from a popular model for co-integration \citep{engle1987co}, called the common trends formulation \citep{stock1988testing}.  As shown in \citet{mcelroy2012signal}, a
co-integrated data process can naturally arise from a co-linear
trend process so long as the noise process (after differencing,
if appropriate) is invertible. If the noise process spectrum
has singularities, then the resulting spectrum of the
data process can have singularities as well, which may be an undesirable feature.

These points are discussed further in the
subsequent text and are here intended to provide motivation
for the need of an invertible stationary model.  Although one
can attempt to reparametrize a VAR or VARMA, in order to guarantee
invertibility, it is quite difficult to achieve aside from
utilizing constrained optimization.  For a VEXP both stability
and invertibility are automatic, while the parameters are
completely unconstrained in $\RR^Q$, where $Q$ is the total
number of parameters.  Moreover, the VEXP class of processes is
arbitrarily dense in the space of stable invertible vector
processes, much in the same way that the EXP process can
approximate a stationary univariate process arbitrarily well.
This approximation can be made arbitrarily accurate, and the
novel algorithms developed herein allow for efficient computation of the cepstral representation.  Without such
algorithms, one could hardly advocate for the VEXP; but with
such tools in hand, both Frequentist and Bayesian analyses become tractable, as
explained briefly below.

Many situations arise in which modeling the data by a stationary (stable) vector model is desirable.  For example, this might
occur if the data had already been made stationary by
differencing, or perhaps by utilizing a common trend structure
for an unobserved component (e.g., see \citet{harvey1990forecasting} or \citet{nyblom2000tests}).  For the VAR class, one would need to
impose restrictions on the parameters to ensure a stable
result, or have recourse to use the Yule-Walker estimates \citep[see the discussion in][]{lutkepohl2007new}, which guarantee stable
outcomes.  However, if a Bayesian treatment is desired, prior elicitation becomes a quagmire, since the implicit restrictions
imply that the parameters must be supported on a complicated
manifold.  The Bayesian treatment for the VARMA class of models is even more challenging.  However, the cepstral approach of the VEXP allows for the entries of each parameter matrix to be any real number,
so that taking independent vague Gaussian priors is a sensible
and coherent choice that guarantees a stable outcome.  

There are several facets of this VEXP process that are
fascinating and non-intuitive.  In particular, because we are studying vector
time series, the algebra that relates the cepstral coefficients
to the Wold coefficients is no longer Abelian, and great care
is needed in working with the matrix exponential.  Background
material as well as our basic VEXP model is provided in Section~\ref{sec:basicVEXP} -- moving from mathematical foundations to the explicit definition and on to algorithmic considerations.  Section~\ref{sec:model}
discusses different aspects associated with modeling using the VEXP and provides details surrounding Bayesian estimation, including stochastic search variable selection.
Section~\ref{sec:SimEx} presents two distinct simulated examples, demonstrating the utility of our approach. Subsequently, two bivariate real-data illustrations involving multi-step ahead forecasting and squared coherence estimation are exhibited in Section~\ref{sec:VEXPMod}.  Section~\ref{sec:Conc} presents concluding discussion.  For convenience of exposition, all proofs and derivations are provided in an Appendix.

\section{The VEXP Model}\label{sec:basicVEXP}

\subsection{Preliminaries: vector time series and the matrix
exponential}

General discussion concerning vector time series is provided in \citet{brockwell1991time}.
Here, we will use $\,^\prime$ for transpose and $\,^*$ for conjugate
transpose of a complex-valued matrix. For a $m$-variate time series, the spectral density
matrix $f$ is a $m \times m$ dimensional matrix function of
frequency $\lambda$, and $f(\lambda)$ is always nonnegative definite, and is often positive
definite (pd). Moreover, the autocovariance function (acf) for a mean-zero process is
defined via $\Gamma_h = \EE [ X_{t+h} X_t^{\prime}]$, and is
related to the spectral density matrix (sdm) via the inverse
Fourier transform (FT):
\begin{equation}
\label{eq:acfInvFT}
  \Gamma_h = \frac{1}{2 \pi} \int_{-\pi}^{\pi} f(\lambda) z^{-h} \,
  d\lambda,
\end{equation}
where $z = \exp({-i \lambda})$.  This integration works
component-wise on each entry of the sdm.  This relationship can
be re-expressed in terms of the FT as follows:
\begin{equation}
 \label{eq:acf2sdm}
 f(\lambda) = \sum_{h = -\infty}^{\infty} \Gamma_h z^h.
\end{equation}
This relation is indicative of a more general Hilbert Space
expansion of spectral matrix functions, where a generic function, $g$,
of frequency can be expanded in terms of the orthonormal basis
$\{ z^h \}$, yielding coefficient matrices given by the inner
product of $g$ with $z^h$.  This is generally true of non-pd
and non-symmetric functions $g$ -- we just compute the basis
expansion for each component function $g_{jk} (\lambda)$ and
then splice the results.

There is a Wold decomposition, or MA($\infty$) representation,
for vector time series, which amounts to a particular form for
the sdm; see \citet{brockwell1991time} for a comprehensive discussion. Let
$\Psi (z) = \sum_{j \geq 0} \Psi_j z^j$ be the causal
representation of the time series (and for identifiability, we
have $\Psi_0 = I$ the identity matrix), such that $X_t = \Psi
(B) \epsilon_t$, for some vector white noise $\{\epsilon_t \}$
with (lag zero) covariance matrix $\underline{\Sigma}$.  Then
the sdm is
\begin{equation*}
 \label{eq:movingAvesdm}
 f (\lambda) = \Psi (z) \underline{\Sigma} \Psi^{\prime}
 (\overline{z}).
\end{equation*}
Given the above definitions, the acf is related to the
Wold filter $\Psi (B)$ by
\begin{equation*}
\label{eq:ma2acf}
 \Gamma_h = \sum_{j \geq 0} \Psi_{j+h} \underline{\Sigma} \Psi_j^{\prime}.
\end{equation*}
Now let us consider a different representation of the sdm that
involves the matrix exponential.  The matrix exponential is
defined in \citet{artin1991}, and is mainly used in the theory of
partial differential equations.  Many of its properties are
elucidated in \citet{chiu1996matrix}.  For any
complex-valued square matrix $A$, the matrix exponential $\exp(A)$
is defined via the Taylor series expansion of $\exp(x)$ evaluated
at $x = A$.  Proposition 8.3 of \citet[][p. 139]{artin1991} guarantees
the convergence.

Since the sdm is pd by assumption, we can always find orthogonal matrices (as
a function of frequency) to diagonalize it.  By
(\ref{eq:acf2sdm}) we know that $f$ is Hermitian, and hence $f
= P A P^{*}$, where $P = P(\lambda)$ is a unitary matrix and $A
= A(\lambda)$ is diagonal with positive entries.  Since $P$ is
unitary, for each $\lambda$ we have $P^* = P^{-1}$, and
therefore the matrix exponential of $f$ is given by \citet[][p. 139]{artin1991} as $\exp(f) = P \exp(A) P^*$, where $\exp(A)$ is a diagonal
matrix consisting of the exponential of the entries of $A$.
In particular, we can write
\[
  f = \exp \{ P\, \log (A) P^* \},
\]
where the diagonal matrix $\log (A)$ consists of the logged
entries of $A$.  That is, $f$ is the matrix exponential of the
Hermitian matrix $P\, \log (A) P^*$, which is no longer pd in
general.  Of course, this matrix function can be expanded in
the Hilbert Space with basis $\{ z^h \}$ as discussed above,
which yields
\[
  P \log (A) P^*  = \sum_{k= -\infty}^{\infty} \Theta_k z^{k}.
\]
This expansion can be calculated by determining each $\Theta_k$
by inverse FT of $P\, \log (A)\,P^*$.  In the case of a univariate
time series, the scalars $\{ \Theta_k \}$ are called cepstral
coefficients \citep{Bloom:1973, holan2004time}; however, in our context they are
matrices.  Hence, we call them cepstral matrices.  Therefore, we
obtain the formal expression
\begin{equation}
 \label{eq:sdmExpansion}
  f  = \exp \left( \sum_{k= -\infty}^{\infty} \Theta_k z^{k}  \right).
\end{equation}
So far we have proceeded generally; that is, a generic sdm of a
covariance stationary vector time series can be written in the
above form, using the matrix exponential.  Now any sdm is
Hermitian, and moreover has the property that $f(-\lambda) =
f^{\prime} (\lambda)$; these properties also hold for $\log(f)$,
and hence we find that the cepstral matrices are real-valued
and satisfy $\Theta_{-k} = \Theta_k^{\prime}$.  Letting $\Theta
(z) = \sum_{k \geq 1} \Theta_k z^k$, (\ref{eq:sdmExpansion})
becomes $f = \exp \{ \Theta_0 + \Theta(z) + \Theta^{\prime}
(\overline{z}) \}$.  Hence, the {\it only} constraints on the
cepstral matrices $\{ \Theta_k \}$ (for $k \geq 0$) are that
they have real-valued entries.  Using these relations, one
might be tempted to write (\ref{eq:sdmExpansion}) as
\begin{equation}
\label{eq:falseExpansion}
  f = \exp \{ \Theta (z) \} \exp \Theta_0 \exp \{ \Theta^{\prime} (\overline{z}) \}.
\end{equation}
In general, this is false, although it is true when all the
exponentiated matrices are commutative -- cf. Proposition 8.9
of \citet[][p. 140]{artin1991}.  In particular, $\exp({A+B}) = \exp(A) \cdot
\exp(B)$ is not true in general, except when $A$ and $B$ commute.
Thus, we must be more careful in the vector case, because the
algebra is no longer Abelian.

\subsection{The VEXP process}
In analogy with the EXP model of \citet{Bloom:1973}, one might
define a VEXP process by truncating (\ref{eq:sdmExpansion}) in
the index $k$. That is, since the cepstral power series can be
written as the sum of $\Theta (z)$, $\Theta_0$, and
$\Theta^{\prime} (\overline{z})$, we could simply truncate the
power series $\Theta(z)$ to a polynomial. However, an
alternative approach would be to utilize the same truncated
polynomials in (\ref{eq:falseExpansion}).  Although both
approaches are identical in the univariate case of the EXP
process, they are actually distinct definitions when the
cepstral power series $\Theta (z)$ are non-Abelian.  In order
to avoid confusions in notation, we will retain the $\Theta$
notation for the cepstral matrices proper, but write $\Psi (z)
= \exp \{ \Omega (z) \}$ for some power series $\Omega(z) =
\sum_{k \geq 1} \Omega_k z^k$. Counter-examples can be
constructed such that $\Omega (z) \neq \Theta (z)$.  However,
as previously noted, when the cepstral matrices are all commutative
(e.g., suppose that each $\Theta_k$ is diagonal) then
(\ref{eq:falseExpansion}) is indeed true, and we can identify
$\Theta (z)$ and $\Omega (z)$.

The commutativity of the cepstral matrices is a strong
condition, and seems to be incompatible with empirical
processes.  If one takes a truncated version of (\ref{eq:sdmExpansion}) as
the definition of the VEXP, then it is necessary to calculate
covariances using (\ref{eq:acfInvFT}) after computing the matrix
exponential.  We have not discovered any convenient algorithm
for expressing the acf directly in terms of the cepstral
coefficients, and it seems likely that no such method exists --
such an algorithm is not even known in the univariate case;
instead the approach popular in the time series literature is
to compute Wold coefficients from cepstral coefficients, and
then construct the acf from the Wold coefficients \citep{mcelroy2012computation}.  This leaves numerical integration, which is quite inconvenient given that the spectral density in
(\ref{eq:sdmExpansion}) can only be calculated for various
frequencies by repeated approximation of the matrix exponential
by its corresponding Taylor series expansion.  Even if this feat were accomplished, the
Wold coefficients would remain unknown, and they are useful for
forecasting and assessing model goodness-of-fit.

Alternatively, if one adopts the second approach of finding the
cepstral representation of the Wold coefficients, then a ready
algorithm relating $\Psi_k$ to the $\{ \Omega_j \}$ is
available and implementable, as described below.  Once the Wold
coefficients are determined to any desired level of accuracy
(i.e., we compute all coefficients up to some cutoff index
$M$), then the acf can be immediately determined.  So long as the
Wold coefficients decay sufficiently rapidly (e.g., their
matrix norm decays geometrically), we have a decent
approximation to the acf that is quickly computed.  This
strategy mirrors the univariate approach \citep{hurvich2002multi-step}, but
is adapted to the non-Abelian algebra intrinsic to multivariate
analysis.  A VEXP model involves truncating $\Omega (z)$ to a
polynomial; it is also proved below that the approximation error can
be made arbitrarily small by taking a sufficiently high order
VEXP model. 

For these reasons, we adopt (\ref{eq:falseExpansion}) as the
basis for our VEXP process, which will be defined as follows.
Let $ {[\Omega]}_1^q (z)$ denote the first $q$ coefficients in
the matrix power series $\Omega (z)$, with
${[\Omega]}_1^{\infty} (z) = \Omega (z)$ as a special case.
Then the order $q$ VEXP process is defined to have the Wold
representation
\begin{equation}
 \label{eq:VEXPwold}
 \Psi (z) = \exp \{  {[\Omega]}_1^q (z) \}.
\end{equation}
The white noise process has covariance matrix
$\underline{\Sigma}$, which we can represent as the matrix
exponential of some real symmetric matrix, say $\Omega_0$, by
Lemma 1 of \citet{chiu1996matrix}.  Hence, we write
$\underline{\Sigma} = \exp({\Omega_0})$. Note that whereas the
coefficients of $\Omega_k$ for $k \geq 1$ are allowed to be any
real numbers, without constraint, the matrix $\Omega_0$ is
symmetric.  Furthermore, since the determinant of
$\underline{\Sigma}$ equals the exponential of the trace of
$\Omega_0$ \citep[Proposition 5.11][p. 286]{artin1991}, any
zero eigenvalues (corresponding to co-linearity of the white
noise) in the covariance matrix can be conceived as
eigenvalues of size $- \infty$ in $\Omega_0$.
This defines the VEXP($q$) process, and its spectral density can be written
\begin{equation}
\label{eq:VEXPspec}
 f(\lambda)  = \exp \{  {[\Omega]}_1^q (z) \} \exp \{ \Omega_0
 \}  \exp \{  {[\Omega^{\prime}]}_1^q (\overline{z}) \}.
\end{equation}
First, we note that the exponential representation in
(\ref{eq:VEXPwold}) when $q = \infty$ is not automatic for
every Wold filter.  Because the VEXP($\infty$) process must be
invertible, by Corollary 8.10 of \citet[][p. 140]{artin1991}, the
inverse of $\exp \{ \Omega (z) \}$ is $\exp \{ - \Omega (z) \}$
-- it follows that $\Psi (z)$ is too, whenever
(\ref{eq:VEXPwold}) holds.  That is, we cannot have $\mbox{det} \Psi (z) = 0$ for $z$ on the unit circle.

Equation (\ref{eq:VEXPwold}) provides a general relationship
between causal Wold power series and the corresponding cepstral
power series.  In the Appendix we provide a more technical
development of the exact conditions necessary for this
relationship to be valid; the key condition on a given Wold
power series is that $\mbox{det} \Psi (z) \neq 0$ for all $z \in D
= \{ z \in \CC : |z| \leq 1 \}$. Conversely, whenever a
cepstral power series $\Omega (z)$ is also well-defined for $z \in
D$, then $\exp \{ \Omega (z) \}$ is well-defined.  In
particular,  ${[ \Omega ]}_1^q (z)$ is always convergent (on
all of $\CC$, not just $D$) so that the VEXP($q$) for $q <
\infty$ is always well-defined, and it follows that the
corresponding Wold power series has non-zero determinant for
all $z \in \CC$.  As a result, a VEXP($q$) is always invertible
(and stable).  Note that, for $k>0$, no special constraints are required on
the coefficients $\Omega_k$ in order to guarantee stability and
invertibility of the process.

As discussed in \citet{brockwell1991time}, and at more length
in \citet{hannan2012statistical}, a VARMA does not have this
property.  In practice, additional conditions on coefficients --
that are quite subtle to enforce -- must be levied in order to
obtain a stable and invertible fit to data.  Another issue with
the VARMA class is the difficulty of identifiability, which is
discussed further below. But first we show that the
approximation of a VEXP($q$) to a VEXP($\infty$) is arbitrarily
close in a mean square sense.
\begin{Proposition}
\label{prop:Cauchy} Consider an invertible time series with
Wold power series $\Psi (B)$ and cesptral power series $\Omega
(B)$, and let $\Psi^{(q)} (z)$ be the Wold power series
corresponding to the truncated cepstral polynomial $ {[
\Omega ]}_1^q (B)$, and write $X_t^{(q)} = \Psi^{(q)} (B)
\epsilon_t$ for each integer $q$. Then the time series
$\{ X_t^{(q)} \}$ forms a Cauchy sequence, and converges in mean
square to $\{ X_t \}$.
\end{Proposition}

This result gives us confidence that any time series with a
causal Wold representation, with $\mbox{det} \Psi (z) \neq 0$ for
all $z \in D$, can be approximated arbitrarily well by a
VEXP($q$) by taking $q$ suitably large.  Of course, the same
can be said of finite order VAR, VMA, or VARMA models, but
such models require nuanced parameter restrictions to achieve
stability, invertibility, and/or identifiability.  Consider the
case of a VMA (the discussion can be extended to VARMA, but
is more complicated due to the possibility of cancelation of
common factors) as described in \citet{lutkepohl2007new}, with polynomial $\Psi (z)$.  Imposing $\mbox{det} \Psi (z) \neq 0$
for all $z \in D$ ensures invertibility and identifiability as
well; only imposing $\mbox{det} \Psi (z) \neq 0$ for all $z$
such that $|z| = 1$ still provides invertibility, but the model
will not be identified.  In contrast, the VEXP($q$) corresponds
to an infinite order VMA with $\mbox{det} \Psi (z) \neq 0$ for all
$z \in D$, so that invertibility is automatic; we can also
prove identifiability -- recall that the parameters of the
VEXP($q$) model are just the individual coefficients of each
cepstral matrix.
\begin{Proposition}
\label{prop:ident} A VEXP($q$) process with $q < \infty$ is
stable, invertible, and identifiable.
\end{Proposition}

When $q = \infty$, the assertion of the proposition is still
true when the cepstral power series $\Omega (z)$ converges for
all $z \in D$, as is evident from the proof.  However, the
VEXP($\infty$) would never be used as a model for real data.

\subsection{Properties of the VEXP process}
In the univariate case, one may differentiate
(\ref{eq:VEXPwold}) with respect to $z$, match coefficients,
and arrive at the recurrence relations given in \citet{pourahmadi1984taylor} and \citet{hurvich2002multi-step}. This produces a recursive relation
involving previously computed Wold coefficients and a finite
number of cepstral coefficients.  Such an approach is
demonstrably false in the multivariate case, because
differentiation of the matrix exponential must allow for the
non-Abelian algebra.  In particular, the derivative of $\exp \{
\Omega (z) \}$ is not equal to $\dot{\Omega} (z) \exp \{ \Omega
(z) \}$, except in the case that the terms in $\Omega (z)$
commute with each other.  Counter-examples to illustrate this
can be constructed for a VEXP(2).  Instead, we can relate the
Wold coefficients to cepstral matrices by expanding the matrix
exponential using a Taylor series and matching corresponding powers
of $z$.  Then straightforward combinatorics provides the following
relationship:
\[
 \Psi_k = \sum_{\ell \geq 1 }  \frac{1}{ \ell !}
 \left( \sum_{ \lambda \vDash k : |\lambda | = \ell}
   \Pi_{j=1}^{\ell} \Omega_{i_j} \right)
\]
for $k \geq 1$.  The symbols in the summation are
defined as follows:  $\lambda \vDash k$ denotes a partition of
the integer $k$ -- actually $\vdash$ is typically used
\citep[][p. 28]{stanley1997enumerative}, but because we care about the order of
the numbers occurring in the partition, we use the notation
$\vDash$ instead.  Also, $|\lambda| = \ell$ says that the number
of elements in the partition is $\ell$. So we sum over all
partitions of the integer $k$ into $\ell$ pieces, say $i_1,
i_2, \ldots, i_{\ell}$ with $\sum_{j=1}^{\ell} i_j = k$.  For
example, the size two partitions of the integer 3 are given by
$(1,2)$ and $(2,1)$, and these must be accounted as distinct
terms in the summation, since $\Omega_1 \Omega_2$ is not equal
to $\Omega_2 \Omega_1$.  Actually, when all the $\Omega_k$
matrices commute with each other, all partitions of a given
size and configuration produce the same result, and the above
formula simplifies.  However, this case is of little
practical interest.  We can also produce a relationship of the
cepstral matrices to the Wold coefficients by expanding the
matrix logarithm and matching powers of $z$:
\[
  \Omega_k = \sum_{\ell \geq 1} \frac{ {(-1)}^{\ell}}{\ell}
\left( \sum_{ \lambda \vDash k : |\lambda | = \ell}
   \Pi_{j=1}^{\ell} \Psi_{i_j} \right).
\]
This is typically of lesser interest in applications.  For
modeling, one posits values for the cepstral matrices, and
determines the Wold coefficients.  Counting the numbers of
partitions is laborious, because the total number of (ordered)
partitions of an integer $k$ is equal to $2^k$.  The first few
Wold coefficients are given by (recall that $\Psi_0 = 1$ by
fiat)
\begin{align*}
 \Psi_1 & = \Omega_1 \\
 \Psi_2 & = \Omega_2 + \Omega_1^2/2 \\
 \Psi_3 & = \Omega_3  + \left( \Omega_1 \Omega_2 +
 \Omega_2 \Omega_1 \right) / 2 + \Omega_1^3/6.
\end{align*}
For higher Wold coefficients, the number of terms quickly grows
out of scope.  Note that for $\Psi_3$ the non-Abelian nature of
the cepstral matrices comes into play, since in general
$\Omega_1 \Omega_2 \neq \Omega_2 \Omega_1$.  However, a simpler
method is available that allows the computer to implicitly
determine the appropriate partitions.  Let $\Upsilon (z) =
\Omega (z) / z$, which is a well-defined power series in $z$.
In the case that $\Omega (z)$ is a degree $q$ matrix
polynomial, then $\Upsilon (z) $ is a degree $q-1$ matrix
polynomial. Denote the $j$-th derivative of a polynomial with
respect to $z$ by the superscript $(j)$.  Then we have the
following useful result.
\begin{Proposition}
\label{prop:Woldformula}   Consider the Wold power series $\Psi
(z)$ such that $\mbox{det}\, \Psi (z) \neq 0$ for $z \in D$, and
cesptral power series $\Omega (z)$. With $\Upsilon (z) = \Omega
(z)/z$, the $k$-th Wold coefficient can be computed by
\begin{equation}
\label{eq:Woldformula}
 \Psi_k = \frac{1}{ k !} \, \sum_{\ell = 1}^k \binom{k}{\ell}
  { \left[ { \Upsilon (z) }^{\ell}  \right] }^{(k- \ell)}
  \vert_{z=0} = \sum_{\ell = 1}^k \frac{1}{ \ell !}
  { \left[ { \Upsilon (z) }^{\ell}  \right] }_{k- \ell}.
 \end{equation}
Letting $\Xi (z) = (\Psi (z) - I)/z$, the $k$-th cepstral coefficient
can be computed by
\begin{equation}
\label{eq:Cepsformula}
 \Omega_k = \frac{1}{ k !} \, \sum_{\ell = 1}^k \binom{k}{\ell} \,
 {(-1)}^{\ell} \, (\ell-1)! \, { \left[ { \Xi (z) }^{\ell}  \right] }^{(k- \ell)}
  \vert_{z=0} = - \, \sum_{\ell = 1}^k \frac{{(-1)}^{\ell}}{ \ell }
  { \left[ { \Xi (z) }^{\ell}  \right] }_{k- \ell}.
 \end{equation}
\end{Proposition}
From an algorithmic standpoint, one is required to generate
powers of the matrix polynomial $\Upsilon (z)$ (or $\Xi (z)$)
and read off the appropriate coefficients.  The product of two
matrix polynomials is easily encoded; the resulting matrix
polynomial has coefficients given by the convolution of the
coefficient matrices, respecting the order of the product.  These programs have been coded in 
R \citep{R:software} and are available upon request; consequently, the
computations for the Wold coefficients are straightforward.

\subsection{Applications of the VEXP model}\label{sec:appVEXP}
In terms of modeling with a VEXP($q$), in the frequentist context, we can proceed with a
higher-order model -- confident by Proposition
\ref{prop:Cauchy} that we can get an arbitrarily accurate
approximation to causal invertible processes -- and then refine
the model by replacing ``small" parameter values with zeroes.  In order
to construct parameter estimates, one proceeds by computing the
acf for any posited parameter values and evaluating the Gaussian
or Whittle likelihood as desired (cf. \citet{brockwell1991time} and \citet{taniguchi2000asymptotic}). Then numerical
optima can be determined using BFGS (a quasi-Newton method also known as a variable metric algorithm) or other methods as desired.

Alternatively, using the exact Gaussian likelihood, we can proceed with estimation using a Bayesian approach.  In this setting, the cepstral model order can be chosen using Bayes factor or by minimizing some previously selected criterion, such as deviance information criterion (DIC) \citep{spiegelhalter2002bayesian} or out-of-sample mean squared prediction error.  Within a given model order, estimation of cepstral matrix entries can proceed using stochastic search variable selection (SSVS) \citep{george1993variable, george1997approaches}.

Having fitted a time series model (see below), one may be
interested in a variety of applications: forecasting, signal
extraction, transfer function modeling, or spectral estimation/plotting,
etc.  The versatility of the VEXP model readily allows us these applications.  Plotting
the Wold filter $\Psi(z)$ for $z = \exp({-i \lambda})$ as a
function of frequency allows us to visualize the transfer
function of the process operating on white noise inputs.
Evaluating (\ref{eq:VEXPspec}) allows plotting of the fitted
spectrum, which becomes arbitrarily accurate as $q$ is
increased.

For forecasting, it is necessary to know either the
autocovariance structure or the Wold coefficients.  \citet{mcelroy2012multi} describes multi-step forecasting
for nonstationary vector time series with a general Wold form,
including integrated VARMA models as special cases.  From that
work, the forecast filter (from an infinite past) for $h$-step
ahead forecasting of a stationary process with invertible Wold
power series $\Psi (z)$ is
\begin{equation}\label{eq:fore}
 \Pi (z) = z^{-h} {[ \Psi ]}_h^{\infty} (z) \, \Psi^{-1} (z).
\end{equation}
Noting that the VAR(1), VMA(1), and VEXP(1) all involve the
same number of unknown coefficients, it is of interest to
compare their $h$-step ahead forecast functions.  As in the
previous subsection, let the VAR(1) be written $\Psi (z) = {(I
- \Phi z )}^{-1}$, whereas the VMA(1) is $\Psi (z) = I + \Psi_1
z$.  Of course the VEXP(1) is $\Psi (z) = \exp (\Omega_1 z
)$, which can be expanded into the Wold form with $\Psi_k =
\Omega_1^k / k!$. Moreover, $z^{-h} {[ \Psi ]}_h^{\infty} (z) =
\sum_{k \geq 0} \Omega_1^{k+h} / (k+h)!$, from which $\Pi (z)$
can be computed by a convolution (note that the matrices
involved are just powers of $\Omega_1$, and hence are Abelian).
The forecast filters for the VAR(1), VMA(1), and VEXP(1) are
then respectively given by
\begin{align*}
 \Pi (z) &  = \Phi^h \\
 \Pi (z) & = 1_{\{ h = 1 \} } \; \sum_{k \geq 0} {(-1)}^k \Psi_1^{k+1} z^k \\
 \Pi (z) & = \sum_{k \geq 0} \sum_{\ell=0}^k \frac{1}{\ell ! (k+h-
 \ell)! } \; \Omega_1^{k+h} z^k.
\end{align*}
Note that the VAR(1) forecast only relies on present data; the
VMA(1) uses past data when $h=1$, but otherwise offers the
pathetic prediction of zero when $h > 1$.  The VEXP(1) uses a
geometrically decaying pattern of weights of past data, like
the VMA(1).  As $h $ increases, all the forecast filters tend
to the zero matrix, essentially dictating that long-run
forecasts are given by the mean for a stationary process.

Generalizing to VAR($q$), VMA($q$), and VEXP($q$), it is difficult to
provide explicit formulas for $\Pi(z)$ (except in the VAR
case), but we know that the VAR($q$) filter utilizes the past $q$
values of the series, whereas the VMA($q$) uses all the data so
long as $h \leq q$; when $h > q$ the filter is zero.  For the
VEXP($q$), a weighted average of all past data is implied.  The
repercussions are that VAR forecasts tend to be based upon
recent activity, even when $h$ is large; VMA and VEXP forecasts
can reach deeper into the past, which may be desirable when $h$
is quite large.  A VARMA forecast filter will have behavior
more like that of a VEXP, but the VEXP can be estimated without
concerns regarding identifiability. 

\section{VEXP Modeling of Vector Time Series}\label{sec:model}
Suppose that we have a sample of size $T$ from a mean-zero $m$-variate
time series $\{ X_t \}$, which we wish to model via a VEXP($q$)
process.  Typically, $q$ is chosen via some model selection criteria or to minimize out-of-sample prediction.  Given $q$, we postulate that the Wold representation can be modeled
via (\ref{eq:VEXPwold}), such that the spectral density can be
expressed as (\ref{eq:VEXPspec}).  To distinguish the model
spectrum from the true spectral density of the process $\{ X_t
\}$, we refer to the latter spectrum as $\widetilde{f}$ and the
former spectrum as $f_{\varpi}$, where $\varpi = \mbox{vec} \{
\Omega_0, \Omega_1, \cdots \Omega_q \}$.  Apart from the mean
of the series (for the non-zero mean case), $\varpi$ completely parametrizes the process.  The case of a non-zero mean is readily handled; e.g., see Section~\ref{subsec:Bayes}.  

\subsection{Likelihood estimation}
The Gaussian likelihood is an appealing objective function,
because maximum likelihood estimates (MLEs) have good statistical
properties, such as consistency and efficiency \citep[see][]{taniguchi2000asymptotic}. Writing $\underline{X} = \mbox{vec} \{
X_1, X_2, \cdots, X_T \}$ and $\underline{\Gamma}_{\varpi}$ for
the $mT $ dimensional covariance matrix of the sample, the log
Gaussian likelihood for a mean-zero sample, scaled by $-2$
(sometimes called the deviance) is
\begin{equation}
 \label{eq:GaussLik}
  \mathcal{D} (\varpi; \underline{X}) = \log \det
  \underline{\Gamma}_{\varpi} + \underline{X}^{\prime} \,
  \underline{\Gamma}_{\varpi}^{-1} \, \underline{X},
\end{equation}
which one seeks to minimize.  Efficient computation of the
quadratic form and log determinant in (\ref{eq:GaussLik}) could
proceed utilizing the multivariate Durbin-Levinson algorithm
described in \citet{brockwell1991time}; knowledge of the
autocovariance function -- once this is computed from the model
with parameter vector $\varpi$ -- determines
$\underline{\Gamma}_{\varpi}$ and thereby the deviance.  Recall
that the entries of each cepstral matrix $\Omega_j$ are
unconstrained (although $\Omega_0$ is symmetric), being allowed to be any real number.  However, in
practice, one might still have recourse to use nonlinear optimization with a
bounding box; although from a theoretical standpoint this is
not necessary.   If any estimated coefficient, or component of $\varpi$, is not
significantly different from zero, the model could be
re-estimated with all such coefficients (or some subset)
constrained to be zero, in order to obtain a more parsimonious
model.  One could also attempt to refine the choice of $q$
utilizing such a procedure.

In order to refine the model of order $q$ or to impose additional sparsity, it is important to have the standard
errors of the parameter estimates.  Assuming conditions
sufficient to guarantee efficiency of the MLEs, the inverse of
the numerical Hessian can be used to approximate the covariance
matrix of the parameter MLEs, appropriately scaled.  Another
way to assess competing models is through the Generalized
Likelihood Ratio (GLR) test described in \citet{taniguchi2000asymptotic}, which amounts to taking the difference of two
deviances (\ref{eq:GaussLik}) for a nesting model and a nested
model; specifically, the deviance for the nested model minus
the deviance for the nesting model.  Then one multiplies by
sample size $T$.  The resulting statistic is always
non-negative and, under the null hypothesis that the nested
model is correct, the asymptotic distribution is $\chi^2$ with
degrees of freedom equal to the difference in the number of
parameters in the two models.  This provides a disciplined
method for selecting the VEXP order $q$, because any VEXP model
is automatically nested within a higher order VEXP model.\footnote{\baselineskip=10pt Note that nesting can be more nuanced than 
this, as setting any entry of any cepstral matrix to zero will produce a nested model.}

There may be interest in using other objective functions. In particular, because there
is some computational cost associated with the inversion of
$\underline{\Gamma}_{\varpi}$, an approximate version of the deviance, known as the Whittle
likelihood, may be preferable for very large sample sizes.  In this case, one replaces
the inverse of $\underline{\Gamma}_{\varpi}$ by the covariance
matrix corresponding to the inverse autocovariances.  The
inverse autocovariances are the autocovariances corresponding to
$f^{-1}_{\varpi} = f_{-\varpi}$; i.e., we obtain the inverse of
the VEXP spectrum by considering the alternative VEXP process
where each coefficient is multiplied by negative one.  This is
true, because ${[\Omega]}_1^q (z)$ commutes with
$-{[\Omega]}_1^q (z)$ for any value of $z$, and $\exp(A) \cdot \exp(B)
= \exp({A+B})$ when matrices $A$ and $B$ commute.  The expression
for the Whittle likelihood given in \citet{taniguchi2000asymptotic} also replaces the log determinant term by the log of the
determinant of the innovation variance matrix, i.e., $\log \det
\exp \{ \Omega_0 \}$.  By (\ref{eq:detId}) of Appendix A.1,
this is equal to the trace of $\Omega_0$.  Therefore, for the mean-zero case, the
deviance of the Whittle likelihood can be written as
\begin{equation}
 \label{eq:WhittleLik}
  \mathcal{W} (\varpi; \underline{X}) = \mbox{tr}( \Omega_0)
  + \underline{X}^{\prime} \,
  \underline{\Gamma}_{-\varpi} \, \underline{X},
\end{equation}
which is to be minimized with respect to $\varpi$ (recall that
$\Omega_0$ consists of the first $m(m+1)/2$ entries of $\varpi$).  While
for some time series models (such as unobserved components
models) the inverse autocovariances are time-consuming to
calculate, they are immediate in the case of a VEXP, given that
we have already computed the autocovariances; that is, the insertion
of a minus sign in the algorithm is all that is needed.  That is, (\ref{eq:WhittleLik}) implies
a speedier algorithm, as no matrix inversion is required.

Although mathematically equal, at first glance, the form of the Whittle likelihood in \citet{taniguchi2000asymptotic} is slightly different from (\ref{eq:WhittleLik}).  This other
expression involves the integral over all frequencies
$\lambda \in [-\pi,\pi]$ of the trace of the periodogram
multiplied by $f^{-1}_{\varpi}$; straightforward algebra yields that
this integral is equal to the quadratic form
$\underline{X}^{\prime} \,   \underline{\Gamma}_{-\varpi} \,
\underline{X}$ in (\ref{eq:WhittleLik}).  The periodogram is
defined to be $I_T (\lambda)$, given by
\[
  I_T (\lambda) = T^{-1} \,  \left( \sum_{t=1}^T X_t e^{-i \lambda
  t } \right) \, \left( \sum_{t=1}^T X_t^{\prime} e^{i \lambda
  t } \right),
\]
which is a rank one matrix.  Some statisticians write the
Whittle likelihood in terms of the periodogram only being
evaluated at Fourier frequencies, which amounts to discretizing
the integral in the exact Whittle likelihood by a Riemann
approximation.  The advantage of doing this further
approximation is that the objective function is then expressed
purely in terms of the periodogram and the model spectral
density, and no calculation of inverse autocovariances is
required at all.  This approximate Whittle likelihood can be
written
\begin{equation}
 \label{eq:WhittleLikT}
  \mathcal{W}_T (\varpi; \underline{X}) = \mbox{tr} (\Omega_0)
  + \frac{1}{2 T} \, \sum_{j= -T}^T \mbox{tr} \{ I_T ( \pi j/T) \,
  f_{-\varpi} (\pi j/T) \}.
\end{equation}
Once the periodogram is computed, the evaluation of
(\ref{eq:WhittleLikT}) is extremely fast: one only needs to
evaluate ${[-\Omega]}_1^q (z)$ for $z$ corresponding to the
Fourier frequencies, and determine the matrix exponential (for
example, via Taylor series directly) and construct
$f_{-\varpi}$ via (\ref{eq:VEXPspec}).  If computation of the
autocovariances is prohibitively expensive (due to large $m$
and/or $q$) then the approximate Whittle likelihood may
be preferable.

\subsection{Bayesian estimation}\label{subsec:Bayes}
For an exact Bayesian analysis, the formal procedure for a Gaussian
VEXP($q$) model requires an exact expression for the 
likelihood.  Although it is possible to pose an approximate Bayesian procedure based on the Whittle (or approximate Whittle) likelihood formulation, in moderate sample sizes our preference is for an exact Bayesian approach.  Nevertheless, in large sample sizes, and/or analyses consisting of a large number of time series, an approximate Bayesian procedure may be preferred.  As previously alluded to, in these cases, implementation using the exact or approximate Whittle specification are both extremely computationally efficient.

Depending on the desired goals of a particular analysis, it is often advantageous to treat the elements of the cepstral matrices as nuisance parameters and average over different model specifications using SSVS \citep{george1993variable, george1997approaches, george2008bayesian}.  This type of Bayesian model averaging \citep{hoeting1999bayesian} implicitly weights the elements of the cepstral matrices through the MCMC sampling algorithm.  This strategy is extremely effective in the context of forecasting \citep{holan2012approach}, where interest resides in a target other than the cepstral matrix elements.  If, instead, the main goal is inferential, then a model corresponding to the posterior mode for each cepstral matrix, from the SSVS algorithm, could be re-estimated or, alternatively, models could be considered without SSVS (e.g., with order selection proceeding through Bayes factor or DIC).

To implement the SSVS algorithm we begin by assuming that the likelihood of $\underline{Y}=(\underline{X}-\mu)'$ is specified as 
\begin{eqnarray*}\label{eq:likeUC}
L(\delta,\varpi\vert\cdot)\propto |\Gamma_{\varpi}|^{-1/2}\exp\left(-\frac{1}{2}\underline{Y}'\Gamma_{\varpi}^{-1}\underline{Y}\right),
\end{eqnarray*}
where $\delta=(\mu_1,\mu_2,\ldots,\mu_m)'$ and $\mu=1_T\otimes\delta$.  Note that non-constant $\mu$ could also easily be considered through straightforward modification of the Markov chain Monte Carlo (MCMC) algorithm (e.g., $\mu$ can be specified in terms of covariates).  We further assume that the elements of $\delta$ and the diagonal elements $\Omega_0$ are in the model with probability one.  For the other elements of $\Omega_j$ ($j=1,\ldots,q$), we specify a SSVS prior based on a mixture of normal distributions.

Let $\gamma_i$, $i=1,\ldots,p=qm^2$, denote a latent zero-one random variable, $V_j=\mbox{vec}(\Omega_j)$ ($j=1,\ldots,q$), and $V=(V_1',\ldots,V_q')'$.  Further, let $V_{jk}$ ($k=1,2,\ldots,m^2$) denote the vectorized elements of $V_{j}$, then $V=(V_{11}, V_{12},\ldots,V_{1m^2},\ldots,V_{qm^2})'=(v_1,\ldots,v_p)'$ and we have 
\begin{eqnarray*}
v_i\vert\gamma_i\sim(1-\gamma_i)N(0, \tau_i^2)+\gamma_iN(0,c_i^2\tau_i^2),
\end{eqnarray*}
with $P(\gamma_i=1)=1-P(\gamma_i=0)=\pi_i$.  In other words,
\begin{eqnarray*}\label{eq:SSVSprior1}
V\vert \gamma\sim N(\mbf{0},D_\gamma R D_\gamma),
\end{eqnarray*}
where $R$ is the prior correlation matrix -- which in our case we assume to be the identity matrix (i.e., $V\vert \gamma\sim N(0,D_\gamma^2)$)
 -- and $D_\gamma\equiv\mbox{diag}(a_1\tau_1,\ldots,a_p\tau_p)$ with $p=qm^2$.  In this case, for $i=1,\ldots, p$, $\gamma_i\stackrel{iid}{\sim}\mbox{Bern}(\pi_i)$, with $\pi_i\equiv 1/2$ and $a_i=1$ if $\gamma_i=0$ and $a_i=c_i$ if $\gamma_i=1$.  
 
Note that $\pi_i$ can be viewed as the prior probability that the $i$-th element of $V$ should be included in the model. Therefore, $\gamma_i =1$ indicates that the $i$-th variable is included in the model. Now, in general, $c_i$, $\tau_i$, and $\pi_i$ are fixed hyperparameters; \citet{george1993variable,george1997approaches} describe various alternatives for their specification.  They suggest that one would like $\tau_i$ to be small so that when $\gamma_i = 0$ it is reasonable to specify an effective prior for the $i$-th element of $V$ that is near zero.  Additionally, one typically wants $c_i$ to be large (greater than 1) so that if $\gamma_i = 1$, then our prior would favor a non-zero value for the $i$-th element of $V$.  
  
To complete the Bayesian model, we need to specify prior distributions for the remaining parameters.   In terms of the mean, we assume that $\delta\sim N(\delta_0,\Sigma_\delta)$, with $\delta_0=(u_1,\ldots,u_m)'$ and $\Sigma_\delta=\mbox{diag}(\sigma^2_{\mu_1},\sigma^2_{\mu_2},\ldots,\sigma^2_{\mu_m})$.  Recall, the diagonal elements of $\Omega_0$ are assumed to be in the model with probability one and, thus, we assume that $\Omega_0\sim N(0,\Sigma_{\Omega_0})$, where $\Sigma_{\Omega_0}=\mbox{diag}(\sigma_1^2,\sigma_2^2,\ldots,\sigma_m^2)$.  Lastly, for ($j=1,\ldots,m$), $\sigma^2_{\mu_j}\sim\mbox{IG}(A_{\mu_j},\,B_{\mu_j})$ and $\sigma^2_{j}\sim\mbox{IG}(A_{j},\,B_{j})$.

In some cases it may be of interest to estimate the VEXP model without conducting Bayesian model averaging through SSVS, as is the case in the example we present involving squared coherence estimation (see Section~\ref{subsec:SqCoh}).   Under this scenario, a prior distribution for the elements of $V_j$ ($j=1,\ldots,q$) needs to be specified.  Letting $V_{jk}$ ($k=1,2,\ldots,m^2$) denote the elements of $V_{j}$, we assume that $V_{jk}\sim N(0,\sigma^2_{jk})$.

In general, regardless of whether a SSVS prior is implemented, the full conditional distributions are not of standard form, with the only exceptions being $\delta$ and the elements of $\Sigma_\delta$ and $\Sigma_{\Omega_0}$.  Consequently, all of the parameters aside from $\delta$ and the elements of $\Sigma_\delta$ and $\Sigma_{\Omega_0}$ can be sampled using a random walk Metroplis-Hastings within Gibbs MCMC sampling algorithm.  Sampling of $\delta$ and the elements of $\Sigma_\delta$ and $\Sigma_{\Omega_0}$ proceeds directly using a Gibbs step, as the full conditionals distribution have a closed form. 

\section{Simulated Examples}\label{sec:SimEx}
To illustrate the utility of the VEXP model we present two distinct simulated examples.  The first example highlights estimation through SSVS, whereas the second example considers estimation without SSVS.  The two simulated examples presented here are designed to demonstrate various aspects associated with the analyses presented in Section~\ref{sec:VEXPMod}.

\subsection{Simulated Example I}\label{subsec:SimExI}
The goal of this example is to illustrate that, given an underlying dependence structure, the modeling approach using SSVS is able to provide shrinkage toward the simulated dependence structure with high probability.  This is especially useful in the context of multi-step ahead forecasting, as presented in Section~\ref{subsec:fore}, where our approach averages over several candidate models with the expectation of improved long-term forecasts.

For illustration, we simulate data based on estimates from a VEXP(4) model, with $T=192$, based on the forecasting example presented in Section~\ref{subsec:fore}.  In particular,  the elements of the cepstral matrices are based on estimated values obtained from a VEXP(4) model applied to the bivariate retail sales forecasting example.  Recalling that $V_j=\mbox{vec}(\Omega_j)$, the exact model used for data generation is given by 
\begin{eqnarray*}
V_0&=&(1.305,0.030,0.030,-2.455)',\\
V_1&=&(0.320,-1.170,0.000,0.250)',\\
V_2&=&(0.120,1.505,0.000,0.210)',\\
V_3&=&(0.135,-0.110,0.000,0.045)',\\
V_4&=&(0.130,-2.560,0.000,0.000)',
\end{eqnarray*}
where the mean of the two time series is set equal to zero (i.e., $\delta=(0,0)'$).  In terms of  prior distributions we assume that $\delta\sim N(\overline{x},\text{diag}(\sigma_{\mu_{1}}^{2},\sigma_{\mu_{2}}^{2}))$, $\mbox{diag}(\Omega_{0}) \sim N(0, \text{diag}(\sigma_{1}^{2},\sigma_{2}^{2})),$ and $ \sigma_{\mu_{1}}^{2},\sigma_{\mu_{2}}^{2}, \sigma_{1}^{2},\sigma_{2}^{2} \sim \text{IG}(A,B)$, where the elements of $\overline{x}$ constitute the estimated sample means for the bivariate time series. In addition, we choose $A=2.1$ and $B=1.1$; i.e., we assume an inverse-gamma distribution with mean and variance both being 1. The prior specification for $\mu$ follows from the fact that, for independent and identically distributed (iid) data, $\overline{x}$ is the maximum likelihood estimate (as well as the asymptotic mean).  Finally, based on a sensitivity analysis (see Section~\ref{subsec:fore}), the hyperparameters for the SSVS were specified as $\tau_i\equiv\tau=.10$ and $c_i\equiv c=10$.  The MCMC sampling algorithm was run for 60,000 iterations with the first 40,000 discarded for burn-in.  Convergence was assessed through visual inspection of the sample chains with no evidence of lack of convergence detected.

Table~\ref{tab:sim4_ssvsa} displays the frequency that a particular cepstral matrix specification appeared in the model throughout the 20,000 post burn-in MCMC iterations.  This table clearly illustrates that the SSVS prior is selecting the data generating model specification with high probability.  Additionally, in cases where competing cepstral matrix specifications are chosen, typically the additional elements selected have parameters estimated relatively close to zero.

In contrast, Table~\ref{tab:sim4_para} presents posterior summaries of the estimated mean and cepstral matrix elements.  Importantly, in all cases, the 95\% credible intervals (CIs) capture the true values, with most intervals relatively narrow.  Although the SSVS is implicitly averaging over several model specifications, the fact that the 95\% CIs capture the true values reinforces the fact that the SSVS is able to recover the correct dependence structure with high probability.

\subsection{Simulated Example II}\label{subsec:SimExII}
The second simulated example considers bivariate spectral estimation and, in particular, estimation of squared coherence, where squared coherence is defined as
\begin{eqnarray}\label{eq:sqcoh}
\rho^2_{X_1\cdot X_2}(\lambda)=\frac{\vert f_{X_1X_2}(\lambda)\vert^2}{f_{X_1X_1}(\lambda)f_{X_2 X_2}(\lambda)}.
\end{eqnarray}
This simulation is designed to behave similar to the bivariate critical radio frequency -- sunspots example presented in Section~\ref{subsec:SqCoh} and uses a VEXP(4), with $T=240$, for illustration.  Additionally, this example does not use SSVS; instead, it demonstrates the VEXP framework in situations where model averaging is not necessarily desired.

The VEXP(4) model used to generate data for this example was based on estimates obtained from the critical radio frequency - sunspots data discussed in Section~\ref{subsec:SqCoh}.  Specifically, the model is given by  
\begin{eqnarray*}
V_0&=&(-0.249,0.211,0.211,-0.023)',\\
V_1&=&(1.343,0.081,0.073,0.803)',\\
V_2&=&(0.261,0.169,-0.109,0.432)',\\
V_3&=&(-0.108,0.160,0.138,0.234)',\\
V_4&=&(0.127,0.080,0.114,0.244)',
\end{eqnarray*}
where the mean of the bivariate time series is set equal to zero (i.e., $\delta=(0,0)'$).  In terms of  prior distributions we assume that  $\delta\sim N(\overline{x},\mbox{diag}(\sigma_{\mu_{1}}^{2},\sigma_{\mu_{2}}^{2}))$,  $\mbox{diag}(\Omega_{0})\sim N(0, \mbox{diag}(\sigma_{1}^2,\sigma_{2}^2))$, $\sigma_{\mu_{1}}^{2},\sigma_{\mu_{2}}^{2}, \sigma_{1}^{2},\sigma_{2}^{2} \sim \mbox{IG}(A,B)$, where $\overline{x}$ is the estimated sample mean for the bivariate time series. Again, we choose $A=2.1$ and $B=1.1$; i.e., we assume an inverse-gamma distribution with mean and variance are both one.  For the off-diagonal element of $\Omega_0$, $\Omega_{0}(1,2)$, and all elements in $\Omega_{j}$ for $j=1,2,3,4$, we assumed a $N(0,10^{2})$ prior distribution. 

\begin{figure}
    \begin{tabular}{c}
     \centerline{
     \includegraphics[height=3.25in, width=5.75in]{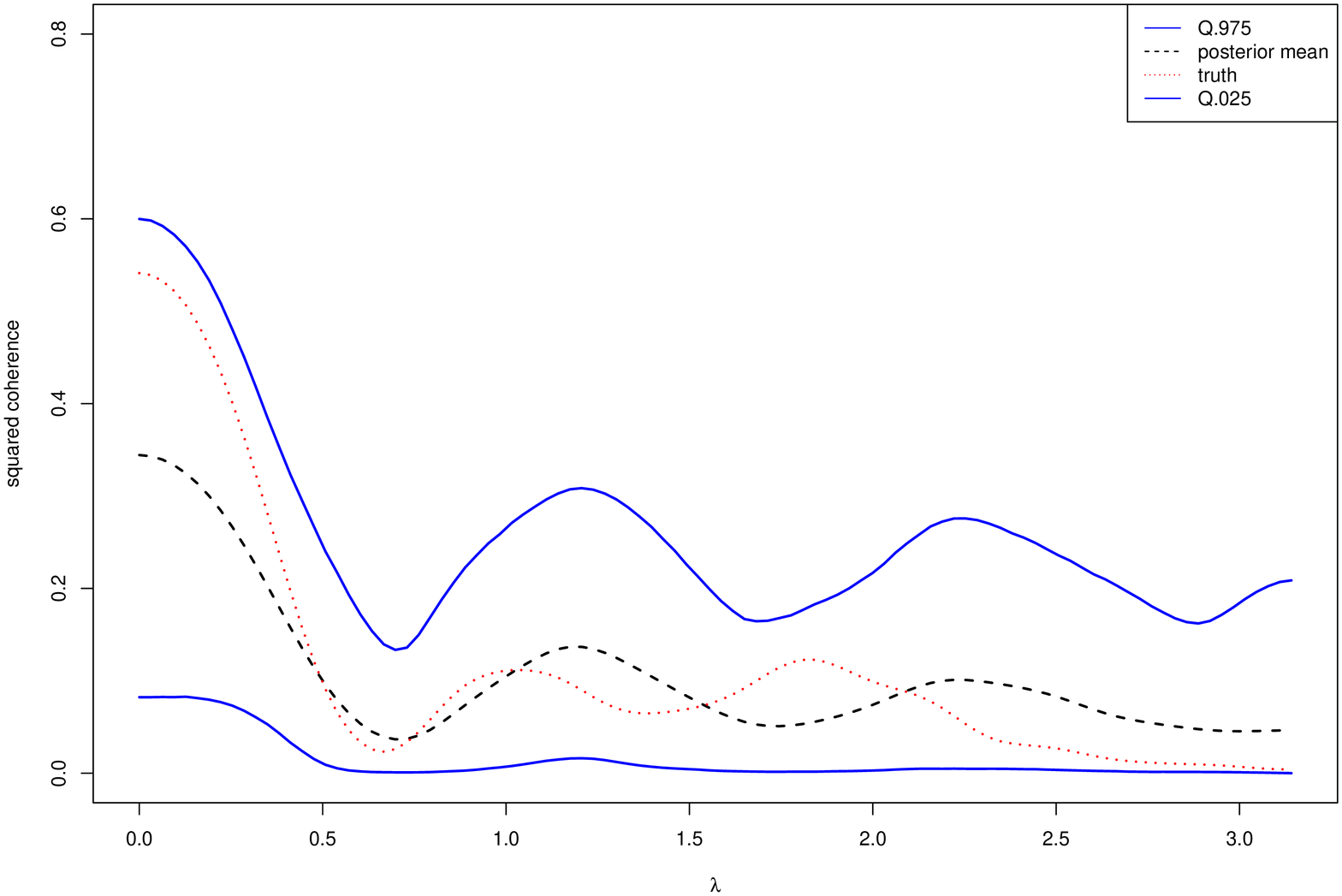}}\\
     (a)\\
    \centerline{
     \includegraphics[height=3.25in, width=6.25in]{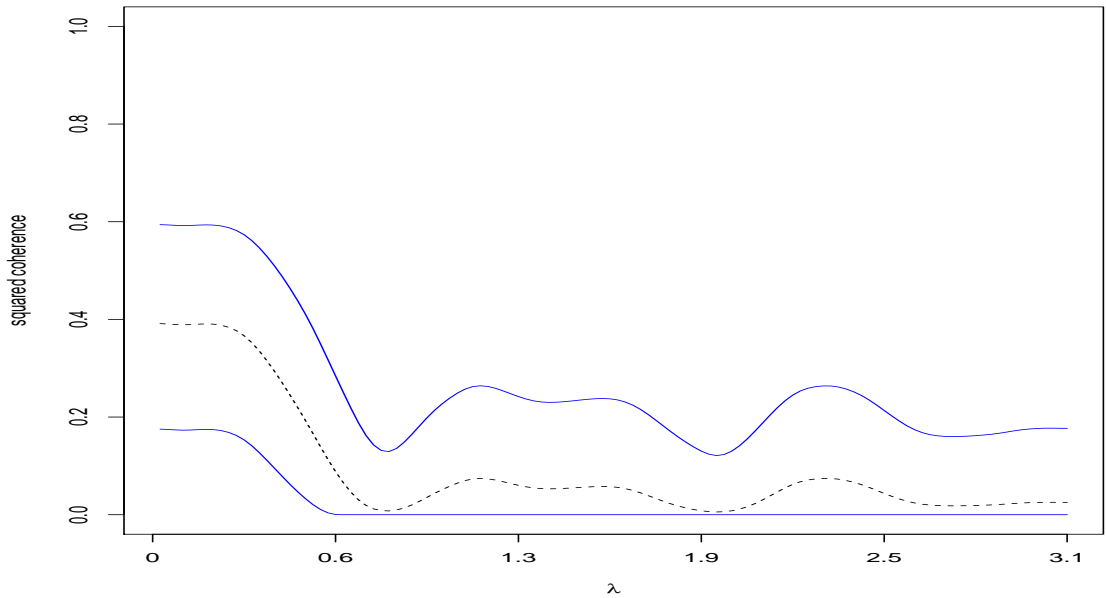}} \\
     (b)   
    \end{tabular}
     \caption{\baselineskip=10pt (a) Comparison of true, pointwise posterior mean, and pointwise posterior 95\% credible intervals of squared coherence for the VEXP(4) model presented in Simulated Example II (Section~\ref{subsec:SimExII}).  Note that the red dotted line and black dashed line denote the truth and posterior mean, respectively.  (b) Empirical squared coherence and pointwise 95\% confidence intervals using modified Daniell window; see Section~\ref{subsec:SimExII}.} 
     \label{fig:SqCohSim}
     \end{figure}

As shown in Table~\ref{tab:sim3_para}, this example clearly demonstrates the ability for our Bayesian estimation procedure to produce reliable results.  In particular, all of the 95\% CIs capture the true values and, in most cases, the intervals are relatively narrow.  Additionally, as depicted in Figure~\ref{fig:SqCohSim}a, the posterior mean squared coherence (obtained as the pointwise mean from the posterior distribution of squared coherence functions) and true squared coherence, as defined by (\ref{eq:sqcoh}), are in close agreement, with the pointwise 95\% CIs relatively narrow away from frequency zero and capturing the true squared coherence.  It is important to note that the deviations between the true and estimated squared coherence in this example are due to the fact that the estimated squared coherence is based on one stochastic realization of the truth, with the Bayesian VEXP(4) estimate agreeing with an empirical estimate obtained through smoothing the multivariate discrete Fourier transform using a modified Daniell window in R (using kernel(``modified.daniell", c(8,8,8)) with taper=.2 in the function spec.pgram).

\section{VEXP Modeling Illustrations}\label{sec:VEXPMod}
To demonstrate the versatility and overall utility of the VEXP modeling framework, we present two real-data examples.  The first example considers multi-step ahead forecasts for a bivariate macroeconomic time series and uses Bayesian model averaging through SSVS as a means of obtaining superior forecasts.  The second example examines the squared coherence between monthly sunspots and critical radio frequencies and does not make use of SSVS.   Instead, the goal of this analysis is to demonstrate the VEXP approach to multivariate spectral (squared coherence) estimation.

\subsection{Multi-step Ahead Forecasting}\label{subsec:fore}
Multi-step ahead forecasting is an area of considerable interest among many scientific disciplines, including atmospheric science and macroeconomics, among others.  One paramount concern when constructing long-lead forecasts is to ensure the model specification is not explosive.  In the context of VAR modeling (or VARMA) this can be facilitated through imposing restrictions on the coefficient matrices to ensure that certain determinants are nonzero.  In contrast, our approach provides an extremely convenient approach to model specification that does not require us to impose any constraints a priori, making estimation exceedingly straightforward within the Bayesian paradigm.

\begin{figure}
     \centerline{
    \includegraphics[height=7.25in, width=7.5in, angle=-90]{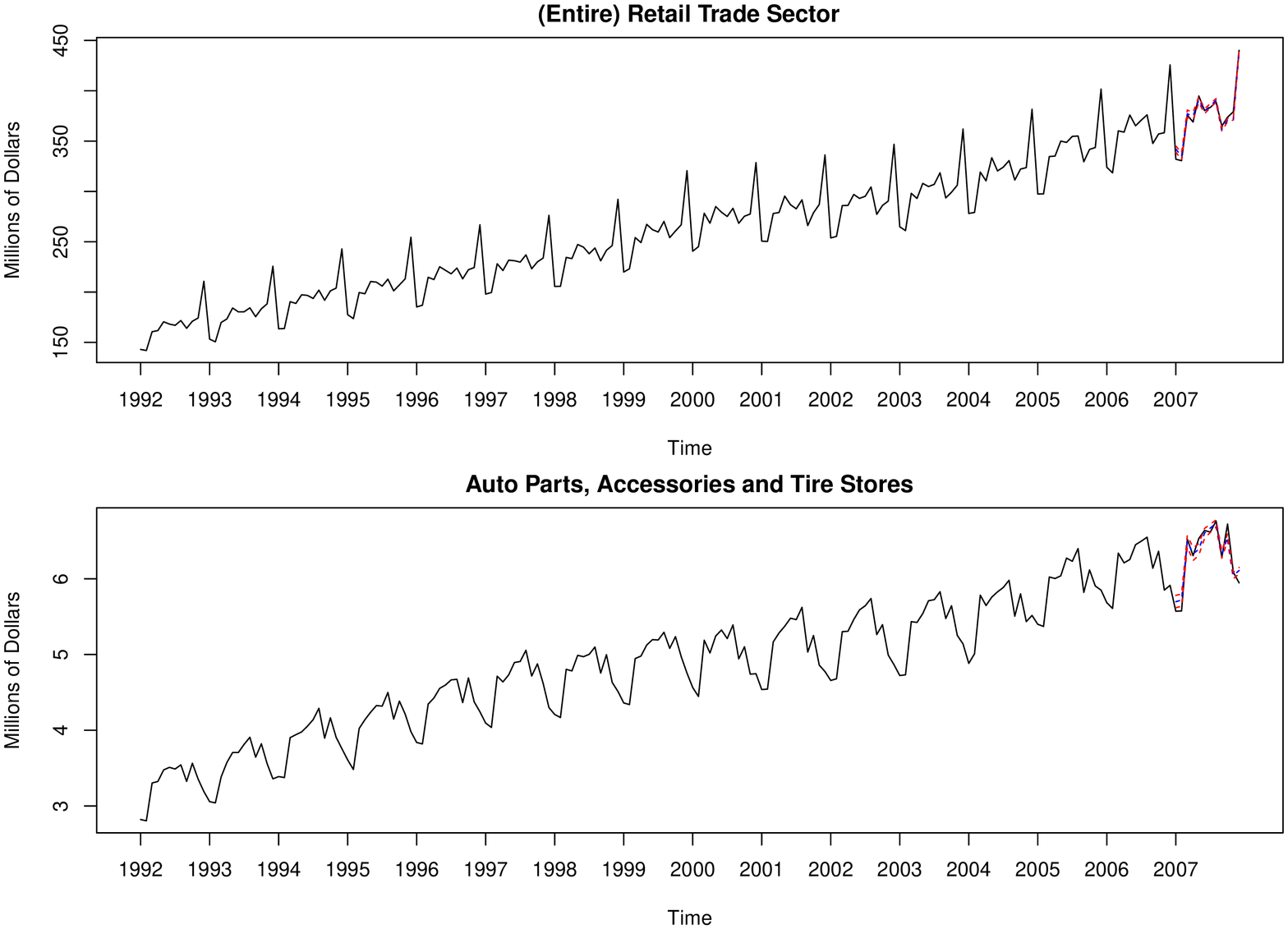}}
     \caption{\baselineskip=10pt Pointwise multi-step ahead (12-steps ahead) forecast plots for the (entire) ``Retail Trade Sector" (RTS) and ``Automotive Parts, Accessories, and Tire Stores" (APATS) series (Section~\ref{subsec:fore}).} 
     \label{fig:foreMCMC}
     \end{figure}

The macroeconomic time series we consider are regression adjusted (i.e., ``Holiday" and ``Trading Day" effects are removed) monthly retail sales time series from the U.S. Census Bureau.  Specifically, we consider a bivariate analysis of (entire) ``Retail Trade Sector" (RTS) and ``Automotive Parts, Accessories, and Tire Stores" (APATS) from January 1992 through December 2007, $T=192$ (Figure~\ref{fig:foreMCMC}).  As previously alluded to, \citet{mcelroy2012multi} describes multi-step forecasting for nonstationary vector time series with a general Wold form.  From that work, the forecast filter (from an infinite past) for $h$-step ahead forecasting for a stationary process with invertible Wold power series $\Psi(z)$ is given by (\ref{eq:fore}).

\begin{figure}
     \centerline{
     \includegraphics[height=5in, width=6in]{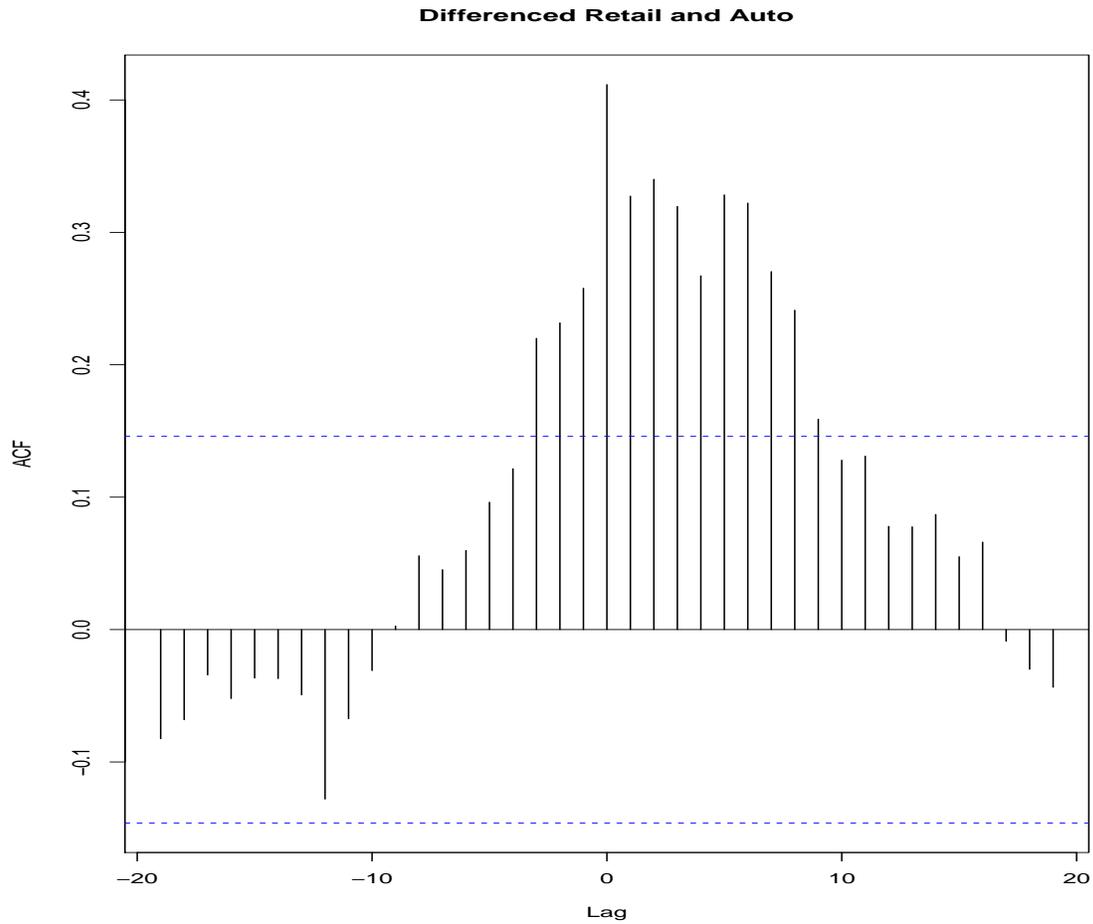}}
     \caption{\baselineskip=10pt Cross-correlation function (ccc) plot for the regression-adjusted and annual-differenced (entire) ``Retail Trade Sector" (RTS) and ``Automotive Parts, Accessories, and Tire Stores" (APATS) time series (Section~\ref{subsec:fore}).} 
     \label{fig:ccf}
     \end{figure}

The bivariate time series considered here are annual-differenced (i.e., the operator $(1-B^{12})$ is applied to the data) prior to estimation using the VEXP model.  From the acf (not shown),  it appears that an AR(1) or low-order ARMA model may be reasonable for each series.  However, the partial autocorrelation function (pacf) is indicative of a lag 12 serial effect, indicating a possible seasonal AR.  As such, the VEXP specification provides a good candidate model.  Additionally, the cross-correlation function (ccf), of the differenced series, indicates that these two series are cross-correlated (Figure~\ref{fig:ccf}).

Prior specification is identical to that of Simulated Example I (Section~\ref{subsec:SimExI}), except in this example we conduct a factorial experiment over various combinations of the SSVS hyperparameters and choose the combination of $c$ and $\tau$ that minimizes the out-of-sample mean squared prediction error (MSPE) over the last 12 values of the series (i.e., minimizes the MSPE associated with the one-step-ahead through twelve-steps-ahead forecasts).  Based on previous forecasting analyses \citep{holan2012approach}, for $q=2,\ldots, 5$, the SSVS hyperparameters considered for this experiment were $\pi_i\equiv\pi= 0.5$ and $(\tau, c)=(0.001,10), (0.001,100)$, $(0.01,10), (0.01,100), (0.1, 10), (0.1, 100)$.  Judging from the overall MSPE, the parameters $q=5$, $\tau=0.1$, and $c=10$ gave  the best performance among the values considered.  Although these hyperparameters could be tuned further, our experience is that such tuning leads to minimal gains in forecasting monthly retail sales.  The SSVS sampler results are based on 60,000 iterations with a 40,000 iteration burn-in.  Convergence is assessed through visual inspection of the sample chains, with no lack of convergence detected.  When conducting out-of-sample forecasting, our $h$-step-ahead forecasts are based on the posterior mean forecast $h$ steps ahead over all iterations of the SSVS MCMC run.  Similar to \citet{holan2012approach}, this forecasting represents a ``model averaging'' over all possible elements of the cepstral matrices, and accounts for their relative importance through the stochastic search procedure.

Using the VEXP model, the MSPE for this example was 21.59 and 0.0112 for the RTS and APATS series, respectively.  In contrast, conducting the same experiment with a VAR(1) model, estimated through OLS, yielded a MSPE of 22.95 and 0.0244 for the RTS and APATS series respectively.  Therefore, relative to the OLS VAR(1) model, the Bayesian VEXP(5) model provides roughly a 6\% decrease in MSPE for the RTS series and a 55\% decrease for the APATS series.  Finally, Figure~\ref{fig:foreMCMC} displays the forecasted values along with their pointwise 95\% CIs and clearly demonstrates the effectiveness of our approach.

\subsection{Modeling of Squared Coherence}\label{subsec:SqCoh}
We consider a monthly bivariate time series of critical radio frequencies and sunspots \citep{newton1988timeslab}.  Specifically, the first series consists of the monthly median noon hour value of the critical radio frequencies (the highest radio frequency that can be used for broadcasting) in Washington D.C. for the period of May 1934 through April 1954.  The second series consists of the total number of monthly sunspots over the same period. 

For this illustration, we consider a VEXP(5) model without SSVS.  Prior specification is identical to that of Simulated Example II (Section~\ref{subsec:SimExII}).  The MCMC sampling algorithm consists of  60,000 iterations with a 40,000 iteration burn-in; i.e., 20,000 iterations are used for inference.  Convergence is assessed through visual inspection of the sample chains, with no lack of convergence detected.  The spectral estimates and squared coherence are obtained by taking the pointwise (by frequency) posterior mean and 95\% CIs.

\begin{figure}
      \begin{tabular}{c}
    \centerline{
     \includegraphics[height=3.5in, width=5.65in]{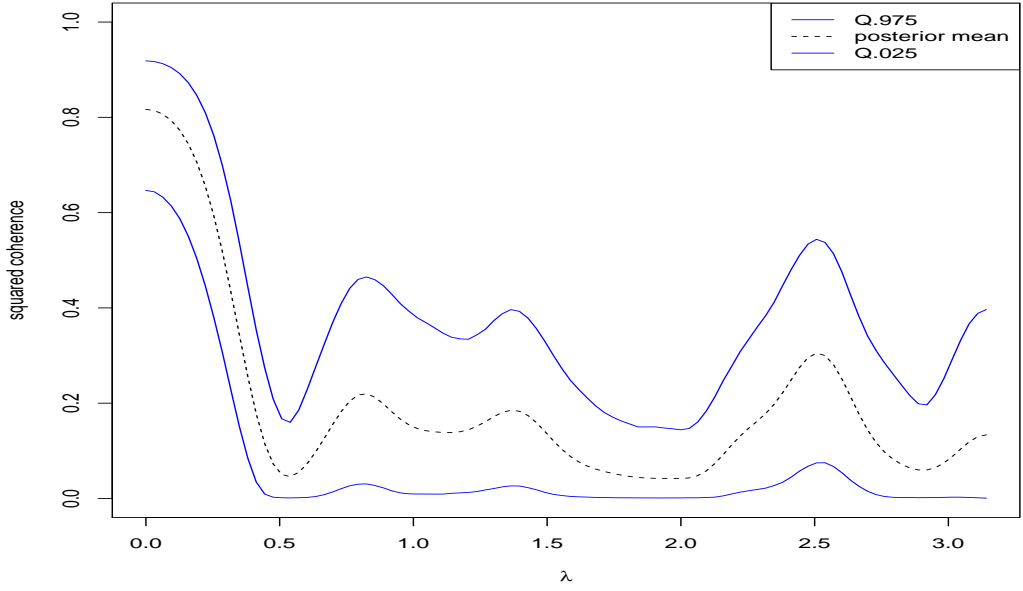}}\\
          (a)\\
    \centerline{
     \includegraphics[height=3.5in, width=5.65in]{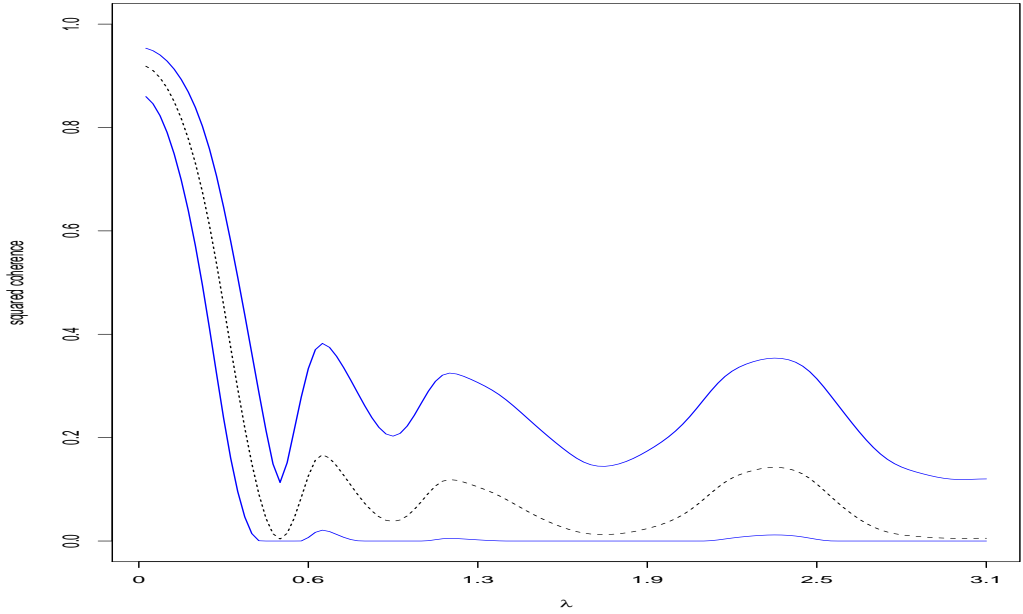}} \\
     (b)   
    \end{tabular}

     \caption{\baselineskip=10pt (a) Pointwise posterior mean squared coherence plot between the critical-radio frequency - sunspots time series corresponding to the period May 1934 through April 1954. (b) Empirical squared coherence and pointwise 95\% confidence intervals using modified Daniell window; see Section~(Section~\ref{subsec:SqCoh}). } 
     \label{fig:sqcoh}
     \end{figure}

The coherence between the two series is illustrated through a plot of the squared coherence (Figure~\ref{fig:sqcoh}a).  From this plot, we see that there is strong coherence at low frequencies (i.e., $\lambda\approx \pi/66$) corresponding to the so-called sunspot cycle ($\approx 11$ years)  Additionally, there is also relatively strong coherence at higher frequencies (i.e.,  $\lambda\approx 2.5$), which may correspond to some sort of (approximately quarterly) seasonal relationship.  These relationships are corroborated through an empirical plot of the squared coherence using a modified Daniell window in R (using kernel(``modified.daniell", c(8,8,8)) with taper=.2 in the function spec.pgram); see Figure~\ref{fig:sqcoh}b.  Discussion regarding these series in the univariate setting can be found in \citet[][p. 194]{newton1988timeslab}.

\section{Conclusion}\label{sec:Conc}
We propose a new class of cepstral models for multivariate time series -- the so-called Vector Exponential Model (VEXP).  Conveniently, this model is cast in the frequency domain and has an unrestricted parameter space.  This is in stark contrast to the VARMA modeling paradigm where one must impose restrictions on the coefficient matrices to ensure that certain determinants are nonzero in order to ensure the model is stationary (or nonexplosive).

We provide theoretical justification for this new class of models and show that this model is dense in the class of short memory time series.  Additionally, for $q<\infty$, we show that the VEXP($q$) process is always stable, invertible, and identifiable.  Importantly, we derive the necessary computational formulas for efficient model implementation and discuss several approaches to estimation, including maximum likelihood and Bayesian estimation.  In fact, one of the primary strengths of the VEXP class of models is that a precise Bayesian treatment proceeds extremely naturally.  In higher dimensional settings (in terms of number of time points and/or series), inversion of the autocovariance matrix in the Gaussian likelihood is computationally expensive.  For these cases, we present an approximate approach based on the Whittle likelihood.

Similar to other multivariate time series models, issues regarding the dimension of the parameter space remain an area of concern.  In the case where the model order and/or the number of series is substantially large, the number of parameters in the model causes difficulty in estimation.  In these cases, further dimension reduction of the cepstral matrices is advantageous and can be achieved through low-rank methods or scientifically motivated parameterizations \citep[e.g., see][]{cressiewikle2011}.  Another practical consideration concerns the number of Wold coefficients, $M$, used for estimation, which needs to be chosen by the practitioner.  In order to guarantee a sufficient approximation, this choice would depend on the underlying dependence structure.  In the short memory cases considered in Sections~\ref{sec:SimEx} and \ref{sec:VEXPMod}, we have taken $M=15$, and found that to be sufficient for our intended purpose.

The methodology is illustrated through simulated examples and through real examples involving multi-step ahead forecasting of bivariate retail trade series from the U.S. Census Bureau and estimation of squared coherence for a bivariate time series of monthly sunspots and critical radio frequencies.  The forecasting example uses Bayesian variable selection in the form of SSVS and thus provides an implicit model averaging.  We demonstrate the superiority of our approach, in terms of MSPE for multi-step-ahead forecasting, relative to an OLS VAR(1) model.  In contrast, the squared coherence example does not impose SSVS and, instead, illustrates various aspects concerning spectral estimation.  Our results corroborate those of previous analyses, while providing a straightforward path to parametric squared coherence estimation.  

Although we highlighted two distinct applications, many other applications of the VEXP model exist.  For example, multivariate long memory modeling and multivariate unobserved component models (e.g., common trends models) using the VEXP framework are two areas of open research.  In summary, any multivariate short memory time series application can be posed using the VEXP framework, thereby providing a rich class of models for multivariate time series.

\section*{Acknowledgments}
This research was partially supported by the U.S. National Science Foundation (NSF) and the U.S. Census Bureau under NSF grant SES-1132031, funded through the NSF-Census Research Network (NCRN) program.

\section*{Appendix}
\begin{appendix}
\Appendix    % This makes the section title start with Appendix!
\renewcommand{\theequation}{A.\arabic{equation}}
\setcounter{equation}{0}

\subsection{Convergence of the Exponential Representation}
Consider equation (\ref{eq:VEXPwold}), which relates $\Psi (z)
$ to the putative $\exp \{ \Omega (z) \}$ (let $q = \infty$ for
this discussion); it is of interest to know when this cepstral representation exists.
For this, we first suppose that $\Psi (z)$ is a causal power series, and
is convergent for $z \in D$.  Then the matrix logarithm of
$\Psi (z)$ is well-defined if and only if (iff) there exists some causal power
series $\Omega (z)$ that converges for $z \in D$ such that
(\ref{eq:VEXPwold}) holds; this is in turn implied by the
condition that ${ \| I - \Psi (z) \| } < 1 $ for $z \in D$, for some matrix norm ${\| \cdot \|}$.   This condition follows from
the expansion of the logarithm:
\[
 \log \Psi (z) = \log \left[ I - \{ I - \Psi (z) \} \right]
  = - \sum_{k \geq 1 } \frac{ {\left\{ I - \Psi (z) \right\} }^k }{k}
 \]
which converges if its matrix norm is finite.  Taking the 
matrix 2-norm, the condition becomes that the eigenvalues of 
$I - \Psi (z)$ have magnitude less than one. This
condition is equivalent to $\mbox{det} \Psi (z) \neq 0$ for all $z
\in D$.

For example, consider the special case that $\Psi (z)$ is a
polynomial, so that $\mbox{det} \Psi (z) = 0$ for some $z \in
\CC$. If any of these zeroes occur within $D$, then $\log\{\Psi
(z)\}$ need not converge, and the cepstral representation is not
valid for $z \in D$.  This is the case of an explosive VMA.  If
instead all the zeroes occur outside of $D$, then $\log\{\Psi
(z)\}$ will converge to $\Omega (z)$, and (\ref{eq:VEXPwold})
holds.

Now suppose that instead we have a well-defined $\Omega (z)$
convergent on $z \in D$, and ask whether $\Psi (z)$ exists in
correspondence.  Because the matrix exponential always
converges (the radius of convergence of the exponential power
series includes all of $\CC$), necessarily $\exp\{\Omega (z)\}$
exists for $z \in D$, and (\ref{eq:VEXPwold}) holds.  This
defines $\Psi (z)$; its determinant will have no zeroes in $D$.
In fact,
\begin{equation}
 \label{eq:detId}
\mbox{det} \Psi (z) = \exp \{ \mbox{tr} \Omega (z) \}
\end{equation}
holds, which shows that the determinant must be non-negative,
and is zero only if $\mbox{tr}\{\Omega (z)\} = - \infty$; but,
this is excluded by the assumption that $\Omega (z)$ converges.
Hence $\mbox{det} \Psi (z) \neq 0$ for all $z \in D$, so that
existence of $\Omega (z)$ automatically ensures invertibility
(and identifiability) of the Wold form.

\subsection{Technical Proofs}

\paragraph{Proof of Proposition \ref{prop:Cauchy}.}
We note that this proof requires the main result of Proposition
\ref{prop:Woldformula}, but in terms of exposition it makes
more sense to state Proposition \ref{prop:Cauchy} first.  To
prove the time series is mean square Cauchy, we take
differences for $q = m+h$ and $q = m$, where $m$ and $h$ are
large integers:
\[
  X_t^{(m+h)} - X_t^{(m)} = \left[ \Psi^{(m+h)} (B) - \Psi^{(m)} (B)
  \right] \epsilon_t.
\]
Here the time index $t$ is immaterial, since we will compute
the covariance matrix of the above vector difference.  The
covariance equals
\[
  \sum_{k \geq 0} \left( \Psi^{(m+h)}_k - \Psi^{(m)}_k \right)
 \underline{\Sigma} { \left( \Psi^{(m+h)}_k - \Psi^{(m)}_k \right)
 }^{\prime}.
\]
In order to show that this matrix tends to zero as $m$ and $h$
grow to infinity, it suffices to examine the sequence $
\Psi^{(m+h)}_k - \Psi^{(m)}_k$, which by Proposition
\ref{prop:Woldformula} can be written as follows.  Let
$\Upsilon (z) = \Omega (z) / z$, where $\Omega (z)$ corresponds
to $\Psi^{(m)} (z) $; but the cepstral representation for the
Wold series $\Psi^{(m+h)} (z)$ equals $\Omega (z)$ plus a
second term $\Xi (z) = \sum_{j=1}^h \Omega_{j+m} z^{j+m}$. Thus
$\{\Omega (z) + \Xi (z)\}/z = \Upsilon (z) + \Pi (z)$, say.  With
these notations, we have
\[
 \Psi^{(m+h)}_k -  \Psi^{(m)}_k = \frac{1}{ k !} \sum_{\ell = 1}^k \binom{k}{\ell}
  \left( { \left[ { \{ \Upsilon (z) + \Pi (z) \} }^{\ell} \right] }^{(k -
  \ell)} - { \left[ { \Upsilon (z)  }^{\ell} \right] }^{(k -
  \ell)} \right) \vert_{z = 0}.
\]
To evaluate this expression further, we must expand the term
$\Upsilon (z) + \Pi (z)$.  Let us denote these matrices, for
any fixed $z$, by $A_0$ and $A_1$ respectively.  Then the
$\ell$-th power of $A_0 + A_1$ can be written as
\begin{equation}
\label{eq:binarySum}
 \sum_{i_1, i_2, \ldots, i_{\ell} \in { \{0,1 \}}^{\ell}}
 \Pi_{j=1}^{\ell} A_{i_j}.
\end{equation}
Here ${ \{ 0,1 \}}^{\ell} $ denotes the space of binary strings
of length $\ell$, and we sum over all such strings.  Note that
we next subtract off ${ \Upsilon (z) }^{\ell}$, which equals
the single summand of (\ref{eq:binarySum}) that corresponds to
the zero string, i.e., $i_j = 0$ for all $1 \leq j \leq \ell$.
Using the linearity of differentiation, we have
\[
 \Psi^{(m+h)}_k -  \Psi^{(m)}_k =
 \frac{1}{ k !} \sum_{\ell = 1}^k \binom{k}{\ell}
 { \left(  \sum_{i_1, i_2, \ldots, i_{\ell} \in { \{0,1 \}}^{\ell}\setminus {\{ 0 \}}^{\ell} }
 \Pi_{j=1}^{\ell} A_{i_j}  \right) }^{(k - \ell)} \vert_{z = 0}.
\]
Note that $\Pi (z)$ occurs in at least one summand of every
term of $  \Psi^{(m+h)}_k -  \Psi^{(m)}_k $.  In taking the
derivatives and evaluating zero, term by term we either produce
zero or an expression involving a coefficient matrix of $\Pi
(z)$.  The smallest coefficient matrix is $\Omega_{m+1}$, so
taking matrix norms it will be sufficient to show that $\|
\Omega_m \| \tends 0$ as $m \tends \infty$.  Since the matrix
logarithm of $\Psi (z)$ is well-defined, we find that
\[
 \sum_{k \geq 1} \Omega_k \Omega_k^{\prime} =
 \frac{1}{2 \pi} \int_{-\pi}^{\pi} \log \Psi (z) \log \Psi^{\prime}
 (\overline{z}) \, d\lambda.
\]
Taking the trace and using the Frobenius norm $\| \cdot \|$, we obtain
$\sum_{k \geq 1} { \| \Omega_k \| }^2$ equals the trace of the
above expression, which in turn is proportional to the integral
of ${ \| \log \Psi (z) \| }^2$; this is finite for all
$\lambda$ because $ \| I - \Psi (z) \| < 1$ (see the discussion in Appendix A.1).  Hence $\|
\Omega_k \|$ is square summable, and in particular the sequence
tends to zero. This establishes that the sequence is Cauchy in
mean square, and hence $X_t^{(q)} \tends X_t^{(\infty)}$ in
mean square, and $X_t = X_t^{(\infty)}$.  $\quad \Box$

\paragraph{Proof of Proposition \ref{prop:ident}.}
Because $q < \infty$, the convergence of ${[\Omega (z) ]}_1^q$
for $z \in D$ is assured, and by results in Appendix A.1 we
know that $\Psi (z)$ is well-defined and invertible.  Because
$\mbox{det} \Psi (z) \neq 0$ for $z \in D$, the Wold filter is
causal as well. The determinant is also finite, which
guarantees stability.  Next, we establish identifiability.

Let $\theta$ denote a parameter vector describing all the
various entries of the cepstral matrices, in some order, and
$f(\lambda; \theta)$ the associate spectral density.
Let $\theta^{(1)}$ and $\theta^{(2)}$ denote two values of the
parameter vector, but with $f( \cdot ; \theta^{(1)}) = f(\cdot
; \theta^{(2)})$.   Writing the spectral density in the form
(\ref{eq:VEXPspec}), we can invert the Wold filters (because
$\mbox{det} \Psi (z) \neq 0 $ holds for $z \in D$) to obtain
\[
\exp \{ \Omega_0^{(1)} \} = \exp \{ - \Omega^{(1)} (z) \} \,
\exp \{  \Omega^{(2)} (z) \} \, \exp \{ \Omega_0^{(2)} \} \,
\exp \{ {[ \Omega^{(2)} (z) ]}^{*} \} \, \exp \{ -{[ \Omega^{(1)} (z) ]}^*
\}.
\]
The composition of the two causal power series $\exp \{ -
\Omega^{(1)} (z) \} $ and $ \exp \{  \Omega^{(2)} (z) \}$ is
another causal power series with leading coefficient of $I$; because the spectral
density has full rank (because $\mbox{tr} (\Omega_0) > -
\infty$), the Wold factorization is unique (see Hannan and
Deistler, 1988).  It follows that we must have $\exp \{
\Omega^{(1)} \} = \exp \{ \Omega^{(2)} \}$ and $\exp \{ -
\Omega^{(1)} (z) \} \, \exp \{  \Omega^{(2)} (z) \} = I$ for
all $z$.  Hence the Wold filters are identically the same;
applying the matrix logarithm reveals that $\Omega^{(1)} (z) =
\Omega^{(2)} (z)$ for all $z  \in D$.  Now we use the
uniqueness of the Fourier basis to learn that each of the
coefficient matrices are the same, and hence $\theta^{(1)} =
\theta^{(2)}$.  This establishes that $\theta \mapsto f(\cdot;
\theta)$ is injective. $\quad \Box$

\paragraph{Proof of Proposition \ref{prop:Woldformula}.}
To prove (\ref{eq:Woldformula}) we begin with the matrix
exponential expansion, which is valid because the series is
invertible:
\[
 \Psi (z)  = \exp \{ z \, \Upsilon (z) \}  = 1 + \sum_{j \geq 1} \frac{ z^j}{ j! } { \Upsilon (z)
 }^j.
\]
Differentiating $k$ times with respect to the complex variable
$z$ yields
\[
  \Psi^{(k)} (z) = \sum_{j \geq 1}  \sum_{\ell = 1}^k
  \binom{k}{\ell} \frac{ \partial^{\ell} }{ \partial z^{\ell}}
    \frac{ z^{j} }{ j! } \cdot \frac{ \partial^{k - \ell} }{
    \partial z^{k - \ell} } { \Upsilon (z) }^j,
\]
where we can use the Abelian product rule because the scalar
quantities commute with the matrix powers ${\Upsilon (z) }^j$.
Interchanging the summations over $j$ and $\ell$, we see that
if we evaluate at $z = 0$ -- which is equivalent to coefficient
matching -- we have $k! \Psi_k$ on the left hand side, but the
right hand side will be zero unless $j = \ell$.  This produces
the first formula of (\ref{eq:Woldformula}), and the second
follows from algebra.  The proof of (\ref{eq:Cepsformula}) is
similar, but using the expansion for the logarithm instead of
the exponential. $\quad \Box$

\end{appendix}

\clearpage\pagebreak\newpage
\baselineskip=14pt \vskip 4mm\noindent
\bibliographystyle{jasa}
\bibliography{SHHTSM_02152014}

\begin{thebibliography}{27}
\newcommand{\enquote}[1]{``#1''}
\expandafter\ifx\csname natexlab\endcsname\relax\def\natexlab#1{#1}\fi

\bibitem[\protect\citename{Artin, }1991]{artin1991}
Artin, M. (1991).
\newblock {\em Algebra\/}.
\newblock Prentice Hall: Englewood Cliffs, New Jersey.

\bibitem[\protect\citename{Bloomfield, }1973]{Bloom:1973}
Bloomfield, P. (1973).
\newblock \enquote{{An exponential model for the spectrum of a scalar time
  series}.}
\newblock {\em Biometrika\/}, 60, 217--226.

\bibitem[\protect\citename{Brockwell and Davis, }1991]{brockwell1991time}
Brockwell, P.~J. and Davis, R.~A. (1991).
\newblock {\em Time {S}eries: {T}heory and {M}ethods\/}.
\newblock Springer, New York.

\bibitem[\protect\citename{Chiu et~al., }1996]{chiu1996matrix}
Chiu, T.~Y., Leonard, T., and Tsui, K.-W. (1996).
\newblock \enquote{The matrix-logarithmic covariance model.}
\newblock {\em Journal of the American Statistical Association\/}, 91, 433,
  198--210.

\bibitem[\protect\citename{Cressie and Wikle, }2011]{cressiewikle2011}
Cressie, N. and Wikle, C. (2011).
\newblock {\em {Statistics for Spatio-Temporal Data}\/}.
\newblock John Wiley and Sons: Hoboken, New Jersey.

\bibitem[\protect\citename{Engle and Granger, }1987]{engle1987co}
Engle, R.~F. and Granger, C.~W. (1987).
\newblock \enquote{Co-integration and error correction: representation,
  estimation, and testing.}
\newblock {\em Econometrica: Journal of the Econometric Society\/},  251--276.

\bibitem[\protect\citename{George and McCulloch, }1993]{george1993variable}
George, E.~I. and McCulloch, R.~E. (1993).
\newblock \enquote{Variable selection via Gibbs sampling.}
\newblock {\em Journal of the American Statistical Association\/}, 88, 423,
  881--889.

\bibitem[\protect\citename{George and McCulloch, }1997]{george1997approaches}
--- (1997).
\newblock \enquote{Approaches for Bayesian variable selection.}
\newblock {\em Statistica Sinica\/}, 7, 2, 339--373.

\bibitem[\protect\citename{George et~al., }2008]{george2008bayesian}
George, E.~I., Sun, D., and Ni, S. (2008).
\newblock \enquote{Bayesian stochastic search for VAR model restrictions.}
\newblock {\em Journal of Econometrics\/}, 142, 1, 553--580.

\bibitem[\protect\citename{Hannan and Deistler, }2012]{hannan2012statistical}
Hannan, E.~J. and Deistler, M. (2012).
\newblock {\em The Statistical Theory of Linear Systems\/}, vol.~70.
\newblock SIAM.

\bibitem[\protect\citename{Harvey, }1990]{harvey1990forecasting}
Harvey, A.~C. (1990).
\newblock {\em {F}orecasting, {S}tructural {T}ime {S}eries {M}odels and the
  {K}alman {F}ilter\/}.
\newblock Cambridge {U}niversity {P}ress, Cambridge.

\bibitem[\protect\citename{Hoeting et~al., }1999]{hoeting1999bayesian}
Hoeting, J.~A., Madigan, D., Raftery, A.~E., and Volinsky, C.~T. (1999).
\newblock \enquote{Bayesian model averaging: A tutorial.}
\newblock {\em Statistical Science\/},  382--401.

\bibitem[\protect\citename{Holan, }2004]{holan2004time}
Holan, S.~H. (2004).
\newblock \enquote{Time series exponential models: Theory and methods.}
\newblock Ph.D. thesis, Texas A\&M University.

\bibitem[\protect\citename{Holan et~al., }2012]{holan2012approach}
Holan, S.~H., Yang, W.-H., Matteson, D.~S., and Wikle, C.~K. (2012).
\newblock \enquote{An approach for identifying and predicting economic
  recessions in real-time using time--frequency functional models.}
\newblock {\em Applied Stochastic Models in Business and Industry\/}, 28, 6,
  485--499.

\bibitem[\protect\citename{Hurvich, }2002]{hurvich2002multi-step}
Hurvich, C.~M. (2002).
\newblock \enquote{Multistep forecasting of long memory series using fractional
  exponential models.}
\newblock {\em International Journal of Forecasting\/}, 18, 2, 167--179.

\bibitem[\protect\citename{L{\"u}tkepohl, }2007]{lutkepohl2007new}
L{\"u}tkepohl, H. (2007).
\newblock {\em New {I}ntroduction to {M}ultiple {T}ime {S}eries {A}nalysis\/}.
\newblock Springer-Verlag: Berlin.

\bibitem[\protect\citename{McElroy and McCracken, }2012]{mcelroy2012multi}
McElroy, T. and McCracken, M.~W. (2012).
\newblock \enquote{Multi-step ahead forecasting of vector time series.}
\newblock {\em Federal Reserve Bank of St. Louis Working Paper Series\/},
  2012-060.

\bibitem[\protect\citename{McElroy and Trimbur, }2012]{mcelroy2012signal}
McElroy, T. and Trimbur, T. (2012).
\newblock \enquote{Signal extraction for nonstationary multivariate time series
  with illustrations for trend inflation.}
\newblock {\em Finance and Economics Discussion Series\/}, 45.

\bibitem[\protect\citename{McElroy and Holan, }2012]{mcelroy2012computation}
McElroy, T.~S. and Holan, S.~H. (2012).
\newblock \enquote{On the computation of autocovariances for generalized
  {G}egenbauer processes.}
\newblock {\em Statistica Sinica\/}, 22, 4, 1661.

\bibitem[\protect\citename{Newton, }1988]{newton1988timeslab}
Newton, H.~J. (1988).
\newblock {\em Timeslab: A Time Series Analysis Laboratory\/}.
\newblock Wadsworth \& Brooks/Cole Publishing, Pacific Grove, CA.

\bibitem[\protect\citename{Nyblom and Harvey, }2000]{nyblom2000tests}
Nyblom, J. and Harvey, A. (2000).
\newblock \enquote{Tests of common stochastic trends.}
\newblock {\em Econometric Theory\/}, 16, 02, 176--199.

\bibitem[\protect\citename{Pourahmadi, }1984]{pourahmadi1984taylor}
Pourahmadi, M. (1984).
\newblock \enquote{Taylor expansion of $\exp \{ \sum_{k=0}^{\infty} a_k z^k \}$
  and some applications.}
\newblock {\em American Mathematical Monthly\/}, 91, 5, 303--307.

\bibitem[\protect\citename{{R Development Core Team}, }2014]{R:software}
{R Development Core Team} (2014).
\newblock {\em R: {A} {L}anguage and {E}nvironment for {S}tatistical
  {C}omputing\/}.
\newblock R Foundation for Statistical Computing, Vienna, Austria.
\newblock {ISBN} 3-900051-07-0.

\bibitem[\protect\citename{Spiegelhalter et~al.,
  }2002]{spiegelhalter2002bayesian}
Spiegelhalter, D.~J., Best, N.~G., Carlin, B.~P., and Van Der~Linde, A. (2002).
\newblock \enquote{Bayesian measures of model complexity and fit.}
\newblock {\em Journal of the Royal Statistical Society: Series B (Statistical
  Methodology)\/}, 64, 4, 583--639.

\bibitem[\protect\citename{Stanley, }1997]{stanley1997enumerative}
Stanley, R.~P. (1997).
\newblock {\em Enumerative {C}ombinatorics\/}.
\newblock Cambridge {U}niversity {P}ress, Cambridge.

\bibitem[\protect\citename{Stock and Watson, }1988]{stock1988testing}
Stock, J.~H. and Watson, M.~W. (1988).
\newblock \enquote{Testing for common trends.}
\newblock {\em Journal of the American Statistical Association\/}, 83, 404,
  1097--1107.

\bibitem[\protect\citename{Taniguchi and Kakizawa,
  }2000]{taniguchi2000asymptotic}
Taniguchi, M. and Kakizawa, Y. (2000).
\newblock {\em Asymptotic Theory of Statistical Inference for Time Series\/}.
\newblock Springer, New York.

\end{thebibliography}

\clearpage\pagebreak\newpage

\begin{table}
\begin{center}
\footnotesize
\begin{tabular}{|c|c|c|}
\hline
Parameters                       &  SSVS &      Freq (out of 20,000)    \\
\hline
$\Omega_{0}(1,2)$         & {\bf 1}          &      11,387   \\
                                            & 0          &        8,613    \\
\hline                                            
$\Omega_{1}$       & {\bf 1101}   &      15,358   \\
                                           & 1111    &        4,528   \\
\hline
$\Omega_{2}$      & {\bf 1101}    &      11,405  \\
                                           & 1100    &         4,500   \\
                                           & 1111    &         2,040  \\
                                           & 1110    &           960  \\
                                           & 0101    &           685  \\
                                           & 0100    &           234  \\
\hline                                         
$\Omega_{3}$      & {\bf 1101}    &         5,005  \\
                                           & 1100    &         4,756  \\
                                           & 0101    &         2,597  \\
                                           & 0100    &         2,668 \\
                                           & 1000    &            739 \\
                                           & 1111    &            733  \\
                                           & 1110    &            725 \\
                                           & 0000    &            503\\
                                           & 0001    &            471 \\
                                           & 0111    &            444 \\
                                           & 0110    &            397 \\
\hline                                         
$\Omega_{4}$                 & 1101    &         7,733  \\
                                           & {\bf 1100}    &         4,476  \\
                                           & 0101    &         3,641  \\
                                           & 0100    &         1,648 \\
                                           & 1111    &         1,107 \\
                                           & 1110    &            627  \\
                                           & 0111    &            533 \\
                                           & 0110    &            235\\
\hline                                         
\end{tabular} 
\end{center}
\caption{\baselineskip=10pt SSVS results for the VEXP(4) model from Simulated Example I (Section~\ref{subsec:SimExI}). Note that only cepstral matrices appearing in the model more than 200 times are detailed in the table and the column labeled SSVS corresponds to an indicator function specifying the elements of $\mbox{vec}(\Omega_j)$ ($j=1,\ldots,4$) appearing in the model.  Finally, note that the bolded entry represents the model structure used to generate the data.}
\label{tab:sim4_ssvsa}
\end{table}

\clearpage\pagebreak\newpage

\begin{table}
\begin{center}
\begin{tabular}{|c|c |c|c|c |c|c|}
\hline
Parameters                       & mean        &   sd            &$Q_{.025}$  &$Q_{.5}$ & $Q_{.975}$  &      True \\
\hline

  $\Omega_{0}(1,1)$ & 1.24532 & 0.11567 & 1.02450 & 1.24352 & 1.47568 & 1.30500 \\ 
  $\Omega_{0}(2,1)$ & -2.47678 & 0.11247 & -2.69252 & -2.47981 & -2.25332 & -2.45500 \\ 
  $\Omega_{0}(2,2)$ & 0.03862 & 0.04419 & -0.01551 & 0.02115 & 0.13369 & 0.03000 \\ 
  $\Omega_{1}(1,1)$ & 0.29931 & 0.07165 & 0.15771 & 0.29951 & 0.43823 & 0.32000 \\ 
  $\Omega_{1}(2,1)$ & -1.28023 & 0.47853 & -2.23111 & -1.28870 & -0.33292 & -1.17000 \\ 
  $\Omega_{1}(1,2)$ & 0.01052 & 0.00875 & -0.00508 & 0.00994 & 0.02985 & 0.00000 \\ 
  $\Omega_{1}(2,2)$ & 0.24884 & 0.06896 & 0.11437 & 0.24841 & 0.38286 & 0.25000 \\ 
  $\Omega_{2}(1,1)$ & 0.16605 & 0.08028 & -0.00144 & 0.16904 & 0.31906 & 0.12000 \\ 
  $\Omega_{2}(2,1)$ & 1.75680 & 0.45747 & 0.88372 & 1.75858 & 2.66037 & 1.50500 \\ 
  $\Omega_{2}(1,2)$ & 0.00590 & 0.00813 & -0.00980 & 0.00581 & 0.02224 & 0.00000 \\ 
  $\Omega_{2}(2,2)$ & 0.08299 & 0.08090 & -0.02128 & 0.07803 & 0.24665 & 0.21000 \\ 
  $\Omega_{3}(1,1)$ & 0.06619 & 0.07615 & -0.02600 & 0.04505 & 0.22631 & 0.13500 \\ 
  $\Omega_{3}(2,1)$ & -0.38706 & 0.48649 & -1.43690 & -0.34742 & 0.46667 & -0.11000 \\ 
  $\Omega_{3}(1,2)$ & 0.00384 & 0.00791 & -0.01112 & 0.00348 & 0.01947 & 0.00000 \\ 
  $\Omega_{3}(2,2)$ & -0.02885 & 0.06065 & -0.17981 & -0.00751 & 0.07065 & 0.04500 \\ 
  $\Omega_{4}(1,1)$ & 0.08438 & 0.08466 & -0.01970 & 0.07374 & 0.25861 & 0.13000 \\ 
  $\Omega_{4}(2,1)$ & -2.79556 & 0.48298 & -3.78661 & -2.79562 & -1.88095 & -2.56000 \\ 
  $\Omega_{4}(1,2)$ & -0.00115 & 0.00773 & -0.01658 & -0.00134 & 0.01391 & 0.00000 \\ 
  $\Omega_{4}(2,2)$ & -0.06388 & 0.07357 & -0.22113 & -0.04491 & 0.03247 & 0.00000 \\ 
  $\mu_{1}$ & -0.10955 & 0.24152 & -0.58511 & -0.11213 & 0.36615 & 0.00000 \\ 
  $\mu_{2}$ & 0.02241 & 0.02709 & -0.03138 & 0.02249 & 0.07616 & 0.00000 \\ 
  $\sigma_{1}^{2}$ & 1.16845 & 1.42409 & 0.28159 & 0.82241 & 4.10003 &  NA \\ 
  $\sigma_{2}^{2}$ & 2.64336 & 6.08424 & 0.62769 & 1.81727 & 8.99748 &  NA \\ 
  $\sigma_{\mu_{1}}^{2}$ & 0.70184 & 0.80633 & 0.16888 & 0.49273 & 2.54956 & NA \\ 
  $\sigma_{\mu_{2}}^{2}$ & 0.69542 & 0.90139 & 0.16607 & 0.48470 & 2.44190 & NA \\ 
   \hline
\end{tabular} 
\end{center}
\caption{\baselineskip=10pt Posterior summary of the VEXP(4) parameters using SSVS for Simulated Example I (Section~\ref{subsec:SimExI}).  Here, ``mean," ``sd," and ``$Q$"denote the posterior mean,  posterior standard deviation, and quantile of the posterior distribution, respectively.}
\label{tab:sim4_para}
\end{table}

\clearpage\pagebreak\newpage

\begin{table}
\begin{center}
\begin{tabular}{|c|c |c|c|c |c|c|}
\hline
Parameters                       & mean        &   sd            &$Q_{.025}$  &$Q_{.5}$ & $Q_{.975}$  &      True \\
\hline
  $\Omega_{0}(1,1)$& -0.14897 & 0.09878 & -0.33487 & -0.15222 & 0.05479 & -0.24900 \\ 
  $\Omega_{0}(2,1)$ & 0.11234 & 0.06381 & -0.01192 & 0.11252 & 0.23769 & 0.21100 \\ 
  $\Omega_{0}(2,2)$ & 0.10710 & 0.10031 & -0.07976 & 0.10539 & 0.31661 & -0.02300 \\ 
  $\Omega_{1}(1,1)$ & 1.31882 & 0.06530 & 1.19197 & 1.31802 & 1.44882 & 1.34300 \\ 
  $\Omega_{1}(2,1)$ & 0.07137 & 0.05760 & -0.04718 & 0.07241 & 0.18201 & 0.08100 \\ 
  $\Omega_{1}(1,2)$ & 0.04971 & 0.07239 & -0.08775 & 0.04805 & 0.19345 & 0.07300 \\ 
  $\Omega_{1}(2,2)$ & 0.78013 & 0.06338 & 0.65332 & 0.78029 & 0.90351 & 0.80300 \\ 
  $\Omega_{2}(1,1)$ & 0.22706 & 0.06532 & 0.10014 & 0.22675 & 0.35491 & 0.26100 \\ 
  $\Omega_{2}(2,1)$ & 0.14242 & 0.05574 & 0.03353 & 0.14276 & 0.25154 & 0.16900 \\ 
  $\Omega_{2}(1,2)$ & -0.05140 & 0.07554 & -0.20083 & -0.05122 & 0.09722 & -0.10900 \\ 
  $\Omega_{2}(2,2)$ & 0.43926 & 0.06320 & 0.31438 & 0.43998 & 0.56269 & 0.43200 \\ 
  $\Omega_{3}(1,1)$ & -0.10610 & 0.06468 & -0.23236 & -0.10613 & 0.02079 & -0.10800 \\ 
  $\Omega_{3}(2,1)$ & 0.08094 & 0.06024 & -0.03730 & 0.08142 & 0.20182 & 0.16000 \\ 
  $\Omega_{3}(1,2)$ & 0.21579 & 0.07655 & 0.06273 & 0.21588 & 0.36484 & 0.13800 \\ 
  $\Omega_{3}(2,2)$ & 0.24103 & 0.06452 & 0.11220 & 0.24061 & 0.36782 & 0.23400 \\ 
  $\Omega_{4}(1,1)$ & 0.15975 & 0.06788 & 0.02469 & 0.15914 & 0.29192 & 0.12700 \\ 
  $\Omega_{4}(2,1)$ & -0.00110 & 0.05933 & -0.11442 & -0.00269 & 0.11792 & 0.08000 \\ 
  $\Omega_{4}(1,2)$ & 0.03664 & 0.07427 & -0.10839 & 0.03684 & 0.18223 & 0.11400 \\ 
  $\Omega_{4}(2,2)$ & 0.20952 & 0.06442 & 0.08318 & 0.20893 & 0.33763 & 0.24400 \\ 
  $\mu_{1}$ & 0.48925 & 0.29485 & -0.09672 & 0.48887 & 1.06666 & 0.00000 \\ 
  $\mu_{2}$ & 0.14156 & 0.33476 & -0.52073 & 0.14177 & 0.79180 & 0.00000 \\ 
  $\sigma_{1}^{2}$ & 0.69913 & 0.81959 & 0.16847 & 0.48853 & 2.50540 &  NA\\ 
  $\sigma_{2}^{2}$ & 0.69215 & 0.83041 & 0.16946 & 0.48956 & 2.41251 &  NA\\ 
  $\sigma_{\mu_{1}}^{2}$ & 0.71532 & 0.81970 & 0.17109 & 0.50502 & 2.53710 & NA \\ 
  $\sigma_{\mu_{2}}^{2}$ & 0.72942 & 0.81193 & 0.17450 & 0.51109 & 2.64821 & NA\\ 
  \hline
\end{tabular} 
\end{center}
\caption{\baselineskip=10pt Posterior summary of the VEXP(4) parameters without SSVS for Simulated Example II (Section~\ref{subsec:SimExII}).  Here, ``mean," ``sd," and ``$Q$"denote the posterior mean,  posterior standard deviation, and quantile of the posterior distribution, respectively.}
\label{tab:sim3_para}
\end{table}
     
\clearpage\pagebreak\newpage

 \clearpage\pagebreak\newpage

\clearpage\pagebreak\newpage

\clearpage\pagebreak\newpage

\end{document}